\documentclass[12pt]{article}
\usepackage{amssymb}
\usepackage{epsfig}
\usepackage{float}
\usepackage{graphicx}
\newcommand{\be}{\begin{equation}}
\newcommand{\ee}{\end{equation}}
\newcommand{\bea}{\begin{eqnarray}}
\newcommand{\eea}{\end{eqnarray}}

\begin{document}

\begin{center}
{\bf NEUTRINOLESS DOUBLE BETA-DECAY}
\end{center}
\begin{center}
S. M. Bilenky
\end{center}

\begin{center}
{\em  Joint Institute for Nuclear Research, Dubna, R-141980,
Russia\\}
\end{center}
\begin{abstract}
The neutrinoless double $\beta$-decay of nuclei is reviewed.
We discuss neutrino mixing and 3$\times$3 PMNS neutrino mixing matrix.
Basic theory of neutrinoless double $\beta$-decay is presented in some details. Results of different calculations of nuclear matrix element are discussed. Experimental situation is considered. The Appendix is dedicated
to E. Majorana (brief biography and his paper in which the theory of Majorana particles is given)

\end{abstract}

\begin{description}
  \item[Introduction]
  \item[Neutrino mixing]
  \item[Seesaw mechanism of the neutrino mass generation]
  \item[Neutrino mixing matrix]
  \item[On neutrino oscillations]
  \item[Basic elements of the  theory of $0\nu\beta\beta$-decay]
   \item[Effective Majorana mass]
  \item[Nuclear Matrix elements]
  \item[Experiments on the search for  $0\nu\beta\beta$-decay]
  \item[Conclusion]
  \item[Appendix. Ettore Majorana ]
\end{description}
\section{Introduction}
Observation  of neutrino oscillations in atmospheric, solar, reactor and accelerator neutrino experiments is one of the most important recent discovery in the particle physics. Small neutrino masses can not be naturally explained by the Standard Higgs mechanism. A new, beyond the Standard Model mechanism of the generation of neutrino masses is required. The most plausible seesaw mechanism of the neutrino mass generation is based on the assumption of the violation of the total lepton number at a large scale and Majorana nature of neutrinos with definite masses.

After it was established that neutrino masses are different from zero, the problem of the nature of neutrinos with definite masses $\nu_{i}$ (Dirac or Majorana?) is the most actual one. Investigation of the neutrino oscillations can not allow to answer this fundamental question. The observation of the neutrinoless double $\beta$-decay ($0\nu\beta\beta$-decay) of some even-even nuclei would be a proof that $\nu_{i}$ are Majorana particles.

The neutrinoless double $\beta$-decay is extremely rare process.
First, this is a process of the second order of the perturbation theory in the Fermi constant. And, second, this process is possible due to helicity-flip. Thus, the matrix element of the process is proportional to the effective Majorana mass $m_{\beta\beta}=\sum_{i}U^{2}_{ei}m_{i}$ ($m_{i}$ is the mass of the neutrino $\nu_{i}$). Smallness of the neutrino masses is an additional reason for smallness of the probability of the $0\nu\beta\beta$-decay.

Very high values for the lower bounds of the half-lives of the $0\nu\beta\beta$-decay of different nuclei were reached in the Heidelberg-Moscow \cite{HMoscow}, IGEX \cite{IGEX}, CUORICINO \cite{Cuoricino} and other experiments. However, in order to reach the values of the half-lives of the $0\nu\beta\beta$-decay which are expected on the basis of the neutrino oscillation data and  the neutrino mass spectrum follows the inverted hierarchy, a new challenging experiments with a sensitivity to $|m_{\beta\beta}|$ about two orders of magnitude better than the today's sensitivity are required. It is expected that such a sensitivity will be reached in several future experiments.

In this review we will consider in some details neutrino mixing. Then we will discuss the standard (Type I) seesaw mechanism of the neutrino mass generation. In the next section we will consider general properties of the neutrino mixing matrix and obtain its standard parametrization. Then we will discuss briefly the present status of neutrino oscillations. In the next section we will present quite detailed derivation of  the matrix element
of the $0\nu\beta\beta$-decay. Then we will consider effective Majorana mass under different assumptions about neutrino mass spectrum. In the next two sections we will discuss the present-day situation with the calculations of nuclear matrix elements of the $0\nu\beta\beta$-decay and experiments on the search for neutrinoless double $\beta$-decay. In the Appendix we will present a short biography of E. Majorana and briefly discuss his 1937 paper  in which the theory of the Majorana particles was developed and a possibility of the existence of such particles was discussed.

For different aspects of the $0\nu\beta\beta$-decay see reviews \cite{DoiKatani,BilPet,Vergados,ElliotVogel,Avignone,VogelPiepke,Vogel}

\section{Neutrino mixing}
We will consider here the neutrinoless double $\beta$-decay under two
general assumptions
\begin{enumerate}
  \item {\em The neutrino interaction is the Standard Model electroweak
  interaction.}
The Lagrangian of the standard charged current (CC) interaction has the form
\begin{equation}\label{CClagrangian}
{\mathcal{L}}^{CC}_{I}(x) = -\frac{g}{2\sqrt{2}}\,j_{\alpha}^{CC}(x)
W^{\alpha}(x) + \mathrm{h.c.}~.
\end{equation}
Here $W^{\alpha}(x)$ is the field of the charged $W^{\pm}$ vector
bosons, $g$ is the constant of the electroweak interaction and
\begin{equation}\label{CCcurrent}
j_{\alpha}^{CC}(x)=2\sum_{l=e,\mu,\tau} \bar \nu_{l L}(x)
\gamma_{\alpha} l_{L}(x)+j_{\alpha}^{h}(x)
\end{equation}
is the sum of the leptonic and hadronic charged current. The hadronic charged current is given by the expression
\begin{equation}\label{CCcurrent1}
j_{\alpha}^{h}(x) = 2(\bar u_{L}(x)\gamma_{\alpha}~d^{\mathrm{mix}}_{L}(x)+
\bar c_{L}(x)\gamma_{\alpha}~s^{\mathrm{mix}}_{L}(x)+
\bar t_{L}(x)\gamma_{\alpha}~b^{\mathrm{mix}}_{L}(x)),
\end{equation}
where
 \begin{equation}\label{quarkmix}
d^{\mathrm{mix}}_{L}(x)=\sum_{q=d,s,b}V_{u q}~q_{L},~
s^{\mathrm{mix}}_{L}(x)=\sum_{q=d,s,b}V_{c q}~q_{L},~
b^{\mathrm{mix}}_{L}(x)=\sum_{q=d,s,b}V_{t q}~q_{L}.
\end{equation}
 In (\ref{quarkmix}) the matrix $V$ is the $3\times 3$ Cabibbo-Kobayashi-Maskawa (CKM) quark mixing matrix \cite{Cab,KobMas}.

The interaction (\ref{CClagrangian}) perfectly describe the
data of numerous experiments on the study of the weak decays, neutrino reactions etc.

\item {\em The neutrino mixing takes place.}

Neutrino fields $\nu_{l L}(x)$ in the leptonic current (\ref{CCcurrent})
are mixed fields
\begin{equation}\label{mixture}
\nu_{lL}(x) = \sum^{3}_{i=1} U_{li} \nu_{iL}(x).
\end{equation}
Here $\nu_{i}(x)$ is the field of neutrino with mass $m_{i}$ and $U$
is the $3\times 3$ Pontecorvo-Maki-Nakagawa-Sakata \cite{BPont,MNS} neutrino
mixing matrix.
\end{enumerate}
The hypothesis of the neutrino mixing was confirmed by the observation of the neutrino oscillations in  experiments with the atmospheric, solar, reactor and accelerator neutrinos. All existing neutrino oscillation data are described if we assume that the number of massive neutrinos is equal to the established number of flavor neutrinos (three).

Quarks are charged particles; the quarks and antiquarks have the same masses and their charges differ in sign. Thus, the quark fields $q(x)$ are complex Dirac fields.

The electric charges of neutrinos are equal to zero. For neutrinos there are two fundamentally different possibilities.
\begin{itemize}
  \item If the total lepton number $L=L_{e}+L_{\mu}+L_{\tau}$ is conserved, neutrino fields $\nu_{i}(x)$ are complex four-component {\em Dirac fields}. In this case neutrinos $\nu_{i}$ and antineutrinos $\bar\nu_{i}$ have the same mass and different lepton numbers ($L(\nu_{i})=-L(\bar\nu_{i})=1$).
  \item If there are no conserved lepton numbers, neutrino fields $\nu_{i}(x)$ are two-component
{\em Majorana fields}. In this case $\nu_{i}\equiv \bar\nu_{i}$.
\end{itemize}
Investigation of the neutrino oscillations does not allow to distinguish these two possibilities \cite{BHP,Doi}. In order to reveal the Majorana nature of $\nu_{i}$ it is necessary to observe processes in which the total lepton number is violated.
{\em Neutrinoless double $\beta$-decay of some nuclei is the only such process the study of which allows to reach the necessary sensitivity.}

The nature of neutrinos with definite masses and the form of the neutrino mixing is determined by  {\em the neutrino mass term of the Lagrangian}. We will consider now possible  mass terms for neutrinos (see \cite{BPet,BGG,ABil}).

A neutrino mass term is the Lorenz-invariant product of the left-handed
and right-handed components of neutrino fields. The three left-handed current fields $\nu_{lL}(x)$, components of $SU(2)$ doublets, must enter into any neutrino mass term. If we assume that three right-handed singlet fields
$\nu_{lR}(x)$ also enter into the Lagrangian  in this case  we can build the following neutrino mass term
\begin{equation}\label{Dmassterm}
\mathcal{L}^{\mathrm{D}}(x)=- \bar\nu_{ L}(x)\,
M^{\mathrm{D}}\,\nu_{ R}(x)
+\rm{h.c.},
\end{equation}
where
\begin{eqnarray}\label{LRculomn}
\nu_{L}=\left(
\begin{array}{c}
  \nu_{e L} \\
  \nu_{\mu L}\\
  \nu_{\tau L}
\end{array}
\right),\quad
\nu_{R}=\left(
\begin{array}{c}
  \nu_{e R} \\
  \nu_{\mu R}\\
  \nu_{\tau R}
\end{array}
\right)
\end{eqnarray}
and $M^{\mathrm{D}}$ is the $3\times 3$ neutrino mass matrix. It is obvious that the total Lagrangian with the neutrino mass term (\ref{Dmassterm})
 is invariant under the global gauge transformations
\begin{equation}\label{globaltransform}
\nu_{ L}(x)\to e^{i\,\Lambda}\nu_{ L}(x),~ \nu_{ R}(x)\to
e^{i\,\Lambda}\nu_{ R}(x),~l_{L,R}(x) \to e^{i\,\Lambda}\,l_{L,R}(x),~ q(x) \to
q(x),
\end{equation}
where $\Lambda$ is an arbitrary constant phase. The invariance under the transformation (\ref{globaltransform}) means that {\em the total lepton number $L$ is conserved}.

The mass term (\ref{Dmassterm}) can be easily diagonalized. For a complex matrix $M^{\mathrm{D}}$ we have
\begin{equation}\label{biunitary}
M^{\mathrm{D}}=U~m~V^{\dag},
\end{equation}
where $U$ and $V$ are unitary $3\times 3$ matrices
and $m$ is a diagonal $3\times 3$ matrix ($m_{ik}=m_{i}~\delta_{ik},~
m_{i}>0$). From (\ref{Dmassterm}) and (\ref{biunitary}) we find
\begin{equation}\label{Dmassterm1}
\mathcal{L}^{\mathrm{D}}(x)=- \bar\nu^{\mathrm{m}}(x)\,
m\,\nu^{\mathrm{m}}(x)=-
   \sum_{i=1}^{3}  m_{i}\,\bar\nu_{i}(x)\,\nu_{i}(x),
\end{equation}
where
\begin{eqnarray}\label{LRculomn1}
 \nu^{\mathrm{m}}_{L}=U^{\dag}\nu_{L}=\left(
\begin{array}{c}
  \nu_{1 L} \\
  \nu_{2 L}\\
  \nu_{3L}
\end{array}
\right),\quad
\nu^{\mathrm{m}}_{R}=V^{\dag}\nu_{L}=\left(
\begin{array}{c}
  \nu_{1 R} \\
  \nu_{2 R}\\
  \nu_{3 R}
\end{array}
\right).
\end{eqnarray}
The expression (\ref{Dmassterm1}) is the sum of standard mass terms for the Dirac fields $\nu_{i}(x)$ with masses $m_{i}$. From (\ref{LRculomn1}) we find that
the flavor fields $\nu_{lL}(x)$ are connected with the left-handed components of the Dirac neutrino fields $\nu_{iL}(x)$ by the mixing relation
\begin{equation}\label{Dmixing}
\nu_{lL }(x)=\sum^{3}_{i=1}U_{li}~\nu_{i L}(x).
\end{equation}
We assumed that  not only  left-handed fields
$\nu_{lL }(x)$ but also the right-handed fields $\nu_{lR }(x)$ enter into the total Lagrangian. In the original Glashow, Weinberg and Salam  papers \cite{Glashow,Weinberg,Salam},  in which the Standard Model was proposed, it was assumed that only $\nu_{lL }(x)$ fields, components of the lepton $SU(2)$ doublets, enter into the Lagrangian. In the seventies
after the success of the  theory of the two-component neutrino it was natural to make this simplest assumption. In such a Standard Model with a $SU(2)$ Higgs doublet neutrinos are massless particles. We can, however, generalize the original SM and to build a model in which neutrino masses and neutrino mixing are generated by the spontaneous violation of the symmetry in the same way as masses and mixing of quarks and leptons. In such a model the neutrino mass term is the Dirac mass term (\ref{Dmassterm}).

We know from experiment that neutrino masses are many orders of magnitude smaller than masses of quarks and leptons. For example, for the particles of the third family
\begin{equation}\label{3gener}
m_{t}\simeq 173~ \mathrm{GeV},~~ m_{b}\simeq 4.2 ~ \mathrm{GeV},~~m_{\tau}\simeq 1.78~ \mathrm{GeV},~~m_{3}\leq 2\cdot 10^{-9}~ \mathrm{GeV}.
\end{equation}
In the framework of the SM there is no natural explanation of such big difference between masses of neutrinos and other fundamental fermions belonging to the same family. It is very implausible that small neutrino masses are generated by the SM Higgs mechanism.

The small Dirac neutrino  masses can be generated, however, in some models
 beyond the SM, for example, in the model
with large extra dimensions \cite{extraD}. In such a model the Newton law at small distances $r$ has the form $F=\frac{1}{M^{2+n}}~\frac{m_{1}m_{2}}{r^{2+n}}$,
where $n$ is the number of the extra dimensions and $M$ is a new scale ($\sim$(1-10) TeV). Dirac neutrino masses in the model with extra dimensions are given by the expression
$$m_{i}\simeq k_{i}v~ \beta.$$
Here $v\simeq$ 250~ GeV is the electroweak scale and $\beta=\frac{M}{M_{P}}\simeq (10^{-15}-10^{-16})$ is a suppression factor ($M_{P}\sim 1.2~10^{19}$~ GeV is the Plank mass).

We will build now a neutrino mass term assuming that fields $\nu_{lL}(x)$
and  $\nu_{lR}(x)$ enter into the Lagrangian and there are no conserved lepton numbers. Let us consider the conjugated fields
\begin{equation}\label{conj}
(\nu_{ L})^{c}=C(\bar\nu_{ L})^{T},\quad(\nu_{
R})^{c}=C(\bar\nu_{ R})^{T},
\end{equation}
where $C$ is the matrix of the charge conjugation which satisfies
the relations
\begin{equation}\label{Cconj}
C\gamma_{\alpha}^{T}C^{-1}=-\gamma_{\alpha},\quad C^{T}=-C.
\end{equation}
It is easy to show that $(\nu_{
L})^{c}$ ($(\nu_{ R})^{c}$) is the right-handed (left-handed) component of the conjugated field.

In fact, for the left-handed and right-handed components we have
\begin{equation}\label{conj1}
\gamma_{5}\,\nu_{L}=-\nu_{L},\quad \gamma_{5}\,\nu_{R}=\nu_{R}.
\end{equation}
From these relations we find
\begin{equation}\label{conj2}
\bar\nu_{L} \,\gamma_{5}=
\bar\nu_{L} ,\quad \bar\nu_{R} \,\gamma_{5}= -\bar\nu_{R}.
\end{equation}
Now, taking into account that $C\gamma_{5}^{T}C^{-1}=\gamma_{5}$, we
have
\begin{equation}\label{conj3}
\gamma_{5}~(\nu_{L})^{c}=(\nu_{L})^{c},\quad
\gamma_{5}~(\nu_{R})^{c}=-(\nu_{R})^{c}.
\end{equation}
From (\ref{conj3}) we conclude that
$(\nu_{
L})^{c}$ and $(\nu_{ R})^{c}$ are the right-handed  and
left-handed components.

The most general neutrino mass term, which can be build from the flavor left-handed fields $\nu_{lL}(x)$ and sterile fields $\nu_{lR}(x)$,\footnote{ Neutrino fields, which do not enter into the Lagrangian of
the standard elecroweak interaction, are called sterile.} has the form
\begin{eqnarray}
\mathcal{L}^{\mathrm{D+M}}= -\frac{1}{2}\,
\bar{\nu_{L}}\,M^{\mathrm{M}}_{L} ( \nu_{L})^{c}-  \bar
\nu_{L}\,M^{\mathrm{D}}\, \nu_{R}
-\frac{1}{2}\,\overline{(\nu_{R})^{c}}\,M^{\mathrm{M}}_{R} \nu_{R}
+\mathrm{h.c.}, \label{DMmassterm}
\end{eqnarray}
where columns $\nu_{L,R}$ are given by (\ref{LRculomn}) and $M_{L,R}^{\mathrm{M}}$ and $M^{\mathrm{D}}$ are nondiagonal complex $3\times 3$  matrices. It is easy to show that $ M_{L,R}^{\mathrm{M}}$ are symmetrical
matrix. In fact, taking into account Fermi-Dirac statistics of the fields $\nu_{ L,R}$, we have
\begin{equation}\label{symmetrical}
\bar\nu_{ L,R}\, M_{L,R}^{\mathrm{M}}C \bar\nu_{ L,R}^{T}=- \bar\nu_{ L,R}\,
(M_{L,R}^{\mathrm{M}})^{T}C^{T} \bar\nu_{ L,R}^{T}=\bar\nu_{ L,R}\,
(M_{L,R}^{\mathrm{M}})^{T}~C \bar\nu_{ L,R}^{T}.
\end{equation}
From this relation we find
\begin{equation}\label{symmetrical1}
M_{L,R}^{\mathrm{M}}=(M_{L,R}^{\mathrm{M}})^{T}.
\end{equation}
It is obvious that the first and the third terms of the expression (\ref{DMmassterm}) are not invariant under the global gauge transformations
$\nu_{ L,R}\to e^{i\Lambda}\nu_{ L,R}$. Thus, in the case of the mass term (\ref{DMmassterm}) the total lepton number $L$ is not conserved.

The first and the third terms of the expression (\ref{DMmassterm}) are called the left-handed and right-handed {\em Majorana mass terms}, respectively. The second term is the Dirac mass term. The mass term $\mathcal{L}^{\mathrm{D+M}}$ is usually called {\em the Dirac and Majorana neutrino mass term} \cite{BilPont,SchechVal}.

 We will show now that in
the case of the mass term (\ref{DMmassterm}) {\em neutrinos with
definite masses are Majorana particles}.

The mass term $\mathcal{L}^{\mathrm{D+M}}$ can be presented in the
following form
\begin{equation}\label{DMmassterm1}
\mathcal{L}^{\mathrm{D+M}}= -\frac{1}{2}\, \bar
n_{L}\,M^{\mathrm{D+M}} (n_{L})^{c} +\mathrm{h.c.}
\end{equation}
Here
\begin{equation}\label{DMmassterm2}
n_{L}={\nu_{L}\choose (\nu_{R})^{c}}
\end{equation}
 and
\begin{eqnarray}
M^{\rm{D+M}}=\left(
\begin{array}{cc}
M^{\mathrm{M}}_{L}&M^{\mathrm{D}}\\
(M^{\mathrm{D}})^{T}&M^{\mathrm{M}}_{R}
\end{array}
\right) \label{DMmatrix}
\end{eqnarray}
is a symmetrical $6\times 6$ matrix.

A symmetrical matrix $M$ can be presented in the form
\begin{equation}\label{Mjdiag1}
M=U\,m\,U^T,
\end{equation}
where $U$ is an unitary matrix and $m$ is a diagonal matrix with positive
diagonal elements.

From (\ref{DMmassterm1})
and (\ref{Mjdiag1})  we have
\begin{eqnarray}\label{DMmassterm2}
\mathcal{L}^{\mathrm{D+M}}&=& -\frac{1}{2}\, \overline{ U^{\dagger}
n_{L}}\,m\,( U^{\dagger} \,n_{L})^{c}
-\frac{1}{2}\, \overline{( U^{\dagger}
n_{L})^{c}}\,m\, U^{\dagger} \,n_{L}\nonumber\\
&=&
-\frac{1}{2}\,\bar\nu^{\mathrm{m}}\,m\,
\nu^{\mathrm{m}}=-\frac{1}{2}\,\sum^{6}_{i=1}\,m_{i}\,\bar\nu_{i}\,\nu_{i}.
\end{eqnarray}
Here
\begin{eqnarray}\label{DMmassterm3}
 \nu^{\mathrm{m}}=U^{\dag} n_{ L}+(U^{\dag} n_{ L})^{c}=\left(
\begin{array}{c}
  \nu_{1 } \\
  \nu_{2 }\\
  \vdots\\
  \nu_{6}
\end{array}
\right).
\end{eqnarray}
From (\ref{DMmassterm2}) and (\ref{DMmassterm3}) we conclude that
\begin{itemize}
  \item The field $\nu_{i}(x)$ ($i=1,2,...6$) is the field of neutrinos with mass $m_{i}$.
  \item The field  $\nu_{i}(x)$ satisfies {\em the Majorana condition}
\begin{equation}\label{Mjcondition}
\nu_{i}(x)=\nu^{c}_{i}(x)=C\bar\nu_{i}^{T}(x).
\end{equation}
\end{itemize}
 Taking into account the unitarity of the  matrix $U$, from (\ref{DMmassterm3}) we find
\begin{equation}\label{DMmixing1}
n_{L} =U\, \nu^{\mathrm{m}}_{L}.
\end{equation}
From (\ref{DMmixing1}) we obtain the following {\em mixing relations in the general Dirac and Majorana case}
\begin{equation}\label{DMmixing2}
\nu_{l L}=\,\sum^{6}_{i=1}U_{l i}\,\nu_{i L},\qquad(\nu_{l R})^{c}=
\,\sum^{6}_{i=1}U_{\bar {l} i}\,\nu_{i L},
\end{equation}
where $U$ is an unitary $6\times6$ mixing matrix and $\nu_{i}$ is the field of the Majorana neutrino with mass $m_{i}$.

Let us discuss the meaning of the Majorana condition (\ref{Mjcondition}). A non hermitian  field $\nu(x)$ can be presented in the following general form
\begin{equation}\label{nufield}
\nu(x)=\int
\frac{1}{(2\pi)^{3/2}}~\frac{1}{\sqrt{2p_{0}}}\left(c_{r}(p)~
u^{r}(p)\,e^{-i\,p\,x} + d^{\dag}_{r}(p)\, u^{r}(-p)
\,e^{i\,p\,x} \right)\,d^{3}p,
\end{equation}
where  $c_{r}(p)$  is the operator of
absorption of neutrino with momentum
$p$ and helicity $r$ and $d^{\dag}_{r}(p)$) is the operator of creation of antineutrino with momentum
$p$ and helicity $r$ and $u^{r}(-p)=C(\bar u^{r}(p))^{T}$.
If the field $\nu(x)$ satisfies the Majorana condition (\ref{Mjcondition})  we find
\begin{equation}\label{Mjcondition1}
c_{r}(p)=d_{r}(p).
\end{equation}
Thus,  if $\nu(x)$ is the Majorana field, {\em
the neutrinos and antineutrinos are identical particles.} In other words the Majorana field is the field of truly neutral particles. There is no notion of particles and antiparticles in the case of the Majorana field.\footnote{In the case of the Majorana field there are no conserved charges which allow to distinguish particles and antiparticles.}

We will finish this section with the following remarks
\begin{enumerate}
\item Dirac and Majorana mass term can be generated only in  theories beyond the SM.
  \item If we assume that only left-handed fields $\nu_{lL}(x)$ enter into the mass term and the lepton number is not conserved, we come the following (Majorana) mass term\cite{GribovPont}
\begin{equation}\label{Mmassterm1}
\mathcal{L}^{\mathrm{M}}= -\frac{1}{2}\,
\bar{\nu_{L}}\,M^{\mathrm{M}}_{L} ( \nu_{L})^{c} +
\mathrm{h.c.},
\end{equation}
where $M^{\mathrm{M}}_{L}$ is $3\times 3$ symmetrical matrix. After the diagonalization,  the mass term (\ref{Mmassterm1}) takes the standard  form
\begin{equation}\label{Mmassterm2}
\mathcal{L}^{\mathrm{M}}=-\frac{1}{2}\,\sum^{3}_{i=1}m_{i}\bar\nu_{i}\nu_{i}
\end{equation}
and we come to the Majorana mixing
\begin{equation}\label{Mmassterm3}
\nu_{l L}=\sum^{3}_{i=1}U_{l i}\,\nu_{i L}.
\end{equation}
Here $U$ is $3\times3$ mixing matrix and $\nu_{i}$ is the Majorana field with the mass $m_{i}$ which satisfies the condition (\ref{Mjcondition}).
Notice that Higgs triplet is needed for the generation of the mass term $\mathcal{L}^{\mathrm{M}}$.

\item
 From the Majorana condition (\ref{Mjcondition}) we have
\begin{equation}\label{Mjcondition2}
 \nu_{iR}=\nu^{c}_{iR}=(\nu_{iL})^{c}.
\end{equation}
Thus, in the case of the Majorana field right-handed and left-handed
components  are connected by the relation
(\ref{Mjcondition2}).  In the case of the Dirac field
right-handed and left-handed components are  independent.
This is the major difference
between Majorana and Dirac fields.

Right-handed components of neutrino fields enter into the Dirac mass term.
If neutrinos are massless there are no mass term in the Lagrangian. This is the reason for the well known theorem \cite{Okubo} which states that
 it is impossible to distinguish massless Dirac and
Majorana neutrinos in the case of left-handed interaction.

\item The Dirac and Majorana mass term   opens
a possibility of the existence of the sterile neutrinos.
If  masses $m_{i}$ are small, in
this case in addition to the mixed flavor left-handed neutrinos $\nu_{e}$, $\nu_{\mu}$
and $\nu_{\tau}$ mixed left-handed antineutrinos $\bar \nu_{l L}$, quanta of mixed right-handed
fields $\nu_{l R}$, must exist. Because right-handed fields do not enter into
the standard CC and NC interactions, $\bar \nu_{l L}$ have no
electroweak interaction. They are called sterile neutrinos. Let us notice that the existed LSND indication in favor of the sterile neutrinos
\cite{LSND} was not confirmed by the MiniBooNE experiment\cite{MiniB}.
\item
In the case of the Dirac and Majorana mass term there are additional sterile right-handed fields $\nu_{lR}$   and
many parameters in the mass matrix. This mass term open a possibility to explain the smallness of the neutrino masses. This (so called seesaw) possibility will be
considered in the next section.
\end{enumerate}

\section{Seesaw mechanism of the neutrino mass generation}
The most popular mechanism of the generation of small neutrino masses is
the seesaw mechanism \cite{seesaw}.
In order to explain the  main idea of this mechanism we consider the simplest case of one generation. The Dirac and Majorana mass term
is given in this case by the  expression
\begin{equation}\label{2DMmassterm}
\mathcal{L}^{\mathrm{D+M}}=-\frac{1}{2}~m_{L} \bar \nu_{L}
(\nu_{L})^{c}-m_{D} \bar \nu_{L} \nu_{R}-\frac{1}{2}~m_{R}
\overline{(\nu_{L})^{c}} \nu_{R} +\mathrm{h.c.}
\end{equation}
We will assume  that $m_{L},m_{D}$ and $m_{R}$ are real
parameters. Let us write Eq. (\ref{2DMmassterm})  in the
matrix form. We have
\begin{equation}\label{2DMjmatrix}
 \mathcal{L}^{\mathrm{D+M}}=-\frac{1}{2}\, \bar n_{L}\,M^{\mathrm{D+M}} (n_{L})^{c}
+\mathrm{h.c.}.
\end{equation}
Here
\begin{eqnarray}\label{2DMjmatrix1}
M^{\rm{D+M}}=\left(
\begin{array}{cc}
m_{L}&m_{D}\\
m_{D}&m_{R}
\end{array}
\right)
\end{eqnarray}
and
\begin{equation}\label{2culomn}
n_{L}={\nu_{L}\choose (\nu_{R})^{c}}~.
\end{equation}
The real symmetrical matrix $M^{\rm{D+M}}$ can be presented in the form
\begin{equation}\label{2diagonalization}
M^{\rm{D+M}}= O\,m'\,O^{T},
\end{equation}
where
\begin{eqnarray}\label{2orthogonal}
O=\left(
\begin{array}{cc}
\cos\theta&\sin\theta\\
-\sin\theta&\cos\theta
\end{array}
\right)
\end{eqnarray}
and $m'_{ik}=m'_{i}\delta_{ik}$, $m'_{i}$ being an eigenvalue of the matrix $M^{\rm{D+M}}$. We have
\begin{equation}\label{2eigenvalues}
m'_{1,2}= \frac{1}{2}\,(m_{R}+m_{L})  \mp
\frac{1}{2}\,\sqrt{(m_{R}-m_{L})^{2} + 4\,m_{D}^{2}}.
\end{equation}
From (\ref{2diagonalization}), (\ref{2orthogonal}) and
(\ref{2eigenvalues}) for the mixing angle $\theta$ we obtain the
following relations
\begin{equation}\label{2mixangle}
\tan 2\,\theta=\frac{2m_{D}}{m_{R}-m_{L}},~~ \cos 2\,\theta=
\frac{m_{R}-m_{L}} {\sqrt{(m_{R}-m_{L})^{2} + 4\,m_{D}^{2}}}~.
\end{equation}
The eigenvalues $m'_{1,2}$ can be positive or negative. Let us
write down
\begin{equation}\label{2eigenvalues1}
 m'_{i}=m_{i}~\eta_{i}~,
\end{equation}
where $m_{i}=|m'_{i}|$ and $\eta_{i}=\pm 1$.

From (\ref{2diagonalization}) and (\ref{2eigenvalues1}) we find
\begin{equation}\label{2diagonalization1}
M^{\rm{D+M}}=U\,m\,U^{T}~,
\end{equation}
where
\begin{equation}\label{2unitarymaqt}
U=O\,\sqrt{\eta}
\end{equation}
 is an unitary matrix.
Using the general results of the previous section, we easily bring the mass term (\ref{2DMmassterm1}) to the standard form
\begin{equation}\label{2DMmassterm1}
\mathcal{L}^{\mathrm{D+M}} =-\frac{1}{2}\bar \nu^{\mathrm{m}}\nu^{\mathrm{m}}
=-\frac{1}{2}\sum_{i=1,2}m_{i}\bar \nu_{i}\nu_{i}.
\end{equation}
Here
\begin{eqnarray}\label{2DMmassterm2}
 \nu^{\mathrm{m}}=U^{\dag} n_{ L}+(U^{\dag} n_{ L})^{c}=\left(
\begin{array}{c}
  \nu_{1 } \\
  \nu_{2 }
  \end{array}
\right),
\end{eqnarray}
$\nu_{i}$ being the Majorana field with the mass $m_{i}$. From (\ref{2DMmassterm2}) we have
\begin{equation}\label{2DMmassterm3}
n_{ L}=U~ \nu_{L}^{\mathrm{m}}.
\end{equation}
Thus, the fields $\nu_{L}$ and $(\nu_{R})^{c}$ are connected with the fields $\nu_{1 L}$
and $\nu_{2 L}$ by the following mixing relations
\begin{eqnarray}
\nu_{L}&=& \cos\theta \sqrt{\eta_{1}}\,\nu_{1 L} + \sin\theta
\sqrt{\eta_{2}}\,\nu_{2 L}
\nonumber\\
(\nu_{R})^{c}&=&- \sin\theta \sqrt{\eta_{1}}\,\nu_{1 L} + \cos\theta
\sqrt{\eta_{2}}\,\nu_{2 L}\label{2mixing}
\end{eqnarray}
Neutrino masses are many orders of magnitude smaller than masses of
leptons and quarks which are generated by the standard Higgs mechanism of
 the electroweak symmetry breaking. This fact is commonly
considered as an evidence in favor of a non-standard  mechanism of  neutrino mass generation. The seesaw mechanism connects
smallness of neutrino masses with the violation of the total lepton
number at a very large scale.

The standard (type I) seesaw mechanism \cite{seesaw} is based on the
following assumptions
\begin{enumerate}
  \item There is no left-handed Majorana mass term in the Lagrangian
  ($m_{L}=0$).
  \item The Dirac mass term is generated by the Higgs
  mechanism ($m_{D}$ is of the order of the mass of a charged
  lepton or quark).
  \item The constant $m_{R}$, which characterize the right-handed Majorana mass term,
the source of the violation of the total lepton number, is much larger than
$m_{D}$:
\begin{equation}\label{seesawinequality}
m_{R}\gg m_{D}
\end{equation}
\end{enumerate}
From (\ref{2eigenvalues}), (\ref{2mixangle}) and
(\ref{seesawinequality}) we have
\begin{equation}\label{masses}
 m_{1}\simeq \frac{m_{D}}{m_{R}}~m_{D}\ll m_{D},~~m_{2}\simeq
 m_{R},~~\tan\theta\simeq \frac{m_{D}}{m_{R}}\ll 1.
\end{equation}
Thus, the seesaw mechanism generates Majorana neutrino  mass $ m_{1}$ which is
much smaller than a Dirac mass of a lepton or quark. As a consequence of the seesaw mechanism a heavy Majorana particle with a mass $m_{2}\simeq m_{R}$ must exist.

Let us consider now the case of the three families.
The  seesaw mixing matrix
has in this case the form
\begin{eqnarray}
M^{\mathrm{seesaw}}=\left(\begin{array}{cc}
0&m_{D}\\
m^{T}_{D}&M_{R}\end{array}\right). \label{3mixingmatrix}
\end{eqnarray}
Here $m_{D}$ is a complex $3\times3$ matrix,  $M_{R}$ is a
symmetrical complex matrix and $m_{D}\ll M_{R}$.

Let us introduce the matrix $M$ by the relation
\begin{equation}\label{3diagonalization}
U^{T}\,M^{\rm{seesaw}}\,U=M,
\end{equation}
where $U$ is an unitary matrix. We will show now the matrix $U$ can be chosen in such a form that $M$ is the block-diagonal matrix.

Notice that in the case of one generation  up to
terms linear in $\frac{m_{D}}{m_{R}}\ll 1$ we have
\begin{eqnarray}
U^{(2)}\simeq\left(\begin{array}{cc}
1&\frac{m_{D}}{m_{R}}\\
-\frac{m_{D}}{m_{R}}& 1\end{array}\right). \label{1mixingmatrix}
\end{eqnarray}
Let us consider the matrix
\begin{eqnarray}
U\simeq\left(\begin{array}{cc}
1&A\\
-A^{\dag}& 1\end{array}\right). \label{1mixingmatrix}
\end{eqnarray}
where $A$ is a $3\times3$ matrix and $A_{ik}\ll 1$. It is easy to
see that up to linear in $A$ terms  $U^{\dag}U\simeq 1$. The non-diagonal
element of the symmetrical matrix $U^{T}\,M^{\rm{seesaw}}\,U$ in the linear over $A$ approximation is equal to
\begin{equation}\label{3nondiagonal}
 m^{T}_{D}-M_{R}~A^{\dag}.
\end{equation}
If we choose
\begin{equation}\label{matrixA}
 A^{\dag}=M^{-1}_{R}~m^{T}_{D}
\end{equation}
the matrix $U^{T}\,M^{\rm{seesaw}}\,U$ takes a block-diagonal form
\begin{eqnarray}
U^{T}\,M^{\rm{seesaw}}\,U\simeq\left(\begin{array}{cc}
-m_{D}~M_{R}^{-1}~m^{T}_{D}&0\\
0& M_{R}\end{array}\right). \label{3massmatrix}
\end{eqnarray}
For the left-handed Majorana neutrino  mass term  from (\ref{3massmatrix}) we
find
\begin{equation}\label{numassterm}
\mathcal{L}^{\mathrm{M}}=-\frac{1}{2}~\bar\nu_{L}M^{\mathrm{M}}_{L}(\nu_{L})^{c}+\mathrm{h.c.}
\end{equation}
where
\begin{equation}\label{}
M^{\mathrm{M}}_{L}= -m_{D}~M_{R}^{-1}~
 m^{T}_{D}
\end{equation}
and $\nu_{L}$ is given by (\ref{LRculomn}).

The Eq. (\ref{numassterm}) is the  mass term for three light
Majorana neutrinos. After the diagonalization of the total mass term
in addition to Majorana neutrino mass term   we will obtain a mass
term for three heavy Majorana particles. Thus, in the case of the Dirac and Majorana
mass term with the matrix (\ref{3mixingmatrix})
in the spectrum of masses there are
\begin{itemize}
\item Three  light Majorana neutrino masses.
\item Three heavy Majorana masses, which are characterized by the scale of
the violation of the total lepton number.
\end{itemize}
These are  general features of the seesaw mechanism. The values of
neutrino masses and mixing angles can be obtained only in the
framework of a concrete model.

Thus, the seesaw mechanism connects smallness of the neutrino masses with
violation of the total lepton number at a large scale.\footnote{Usually it is assumed that this scale is about
$(10^{15}-10^{16})\, \rm{GeV}$.}
 The observation of the neutrinoless double
$\beta$-decay would be an evidence in favor of this mechanism.
Let us notice that the existence of heavy Majorana particles, seesaw partners  of neutrinos, could allow to explain the baryon asymmetry of the
Universe (see \cite{Buch}).

\section{Neutrino mixing matrix}

 In this section we will consider the general properties of the unitary $3\times3$ Dirac (or Majorana) mixing matrix.

An unitary $n\times
n $  matrix $U$ is characterized by $n^{2}$ real
parameters.\footnote{In fact, it can be presented in the form
$U=e^{iH}$, where $H$ is the hermitian matrix. The hermitian matrix is characterized  by $n
+2~(\frac{n^{2}-n}{2})=n^{2}$ real
parameters.} The number of the angles which characterize the unitary $n\times n $
matrix coincides with the number of parameters which characterize
a real orthogonal $n\times n $ matrix $O$ ($O^{T}O=1$). Thus, for the number of the angles we have\footnote{The orthogonal matrix $O$ can be presented in the form $O=e^{A}$, where
$A^{T}=-A$ . Diagonal elements of the matrix $A$ are equal to zero.
The number of the real non-diagonal elements is equal to
$\frac{n(n-1)}{2}$.}
\begin{equation}\label{nangles}
n_{\mathrm{ang}}=\frac{n(n-1)}{2}.
\end{equation}
Other parameters of the matrix $U$ are phases. The number of phases
is equal to
\begin{equation}\label{nphases}
n_{\mathrm{ph}}=n^{2}-\frac{n(n-1)}{2}=\frac{n(n+1)}{2}.
\end{equation}
The number of {\em physical phases} in the neutrino mixing matrix is smaller than $n_{\mathrm{ph}}$. The neutrino mixing matrix enters into the charged current. Let us consider first the case the of Dirac neutrinos $\nu_{i}$.
Because phases of the Dirac fields $l_{L}(x)$ and $\nu_{iL}(x)$ are arbitrary, the matrices $U$ and
\begin{equation}\label{physU}
U'=S^{\dag}(\beta)~U~S(\alpha)
\end{equation}
 are equivalent. Here $S_{ll'}(\beta)=e^{i\beta_{l}}\delta_{ll'}$,
$S_{ik}(\alpha)=e^{i\alpha_{i}}\delta_{ik}$ and  $\beta_{l}$,
$\alpha_{i}$ are real, arbitrary phases.

We can use this freedom in order to exclude $(2n-1)$ phases from the matrix $U$.\footnote{ We can always make one element of the matrix
$S(\alpha)$ (or $S(\beta)$) equal to one. In fact, let us present the matrix $S(\alpha)$ in the form $S(\alpha)=e^{i\alpha_{n}}~S(\bar\alpha)$, where $\bar\alpha_{i}=\alpha_{i}-\alpha_{n}$. The phase factor $e^{i\alpha_{n}}$ can be, obviously, included into $S^{\dag}(\beta)$. We have in this case
$S^{\dag}(\beta)~e^{i\alpha_{n}}=S^{\dag}(\bar\beta)$, where $\bar\beta_{l}=\beta_{l}-\alpha_{n}$.} Thus, in the case of the Dirac neutrinos the number of the physical phases in the mixing matrix $U$
is equal to
\begin{equation}\label{physphase}
 \bar n_{\mathrm{ph}}=\frac{n(n+1)}{2}-(2n-1)=\frac{(n-1)(n-2)}{2}.
\end{equation}
In
the case of the mixing of the three Dirac neutrinos the
mixing matrix is characterized by three mixing angles and one
phase.

Let us consider now the case of the Majorana neutrinos $\nu_{i}$.
The Majorana condition
\begin{equation}\label{Mjc}
\nu^{c}_{i}(x)=\nu_{i}(x)
\end{equation}
does not allow to include arbitrary phases into  the Majorana fields.
For the number of the physical phases we have in the Majorana case \cite{BHP,Doi}
\begin{equation}\label{NMjphase}
\bar n^{M}_{\mathrm{ph}}=\frac{n(n+1)}{2} -n=\frac{n(n-1)}{2}.
\end{equation}
Thus, in
the case of the three Majorana neutrinos the
mixing matrix is characterized by three mixing angles and
three phases.

We will obtain now constraints on the neutrino mixing matrix which follow from the condition of the $CP$ invariance in the lepton sector. Let us consider first the Dirac neutrinos $\nu_{i}$. The condition of the $CP$ invariance
in the lepton sector has the form
\begin{equation}\label{CPinv}
V_{CP}~\mathcal{L}_{I}^{CC}(x)~V^{-1}_{CP}=\mathcal{L}_{I}^{CC}(x').
\end{equation}
Here $V_{CP}$ is the operator of the $CP$  conjugation, $x'=(x^{0},-\vec{x})$
and
\begin{equation}\label{CPCCL}
\mathcal{L}_{I}^{CC}(x)=-\frac{g}{\sqrt{2}}\sum_{l,i}\bar
l_{L}(x)~\gamma_{\alpha}~U_{li}~\nu_{iL}(x)~W^{\alpha\dag}
-\frac{g}{\sqrt{2}}\sum_{l,i}\bar
\nu_{iL}(x)~\gamma_{\alpha}~U^{*}_{li}~l_{L}(x)~W^{\alpha}
\end{equation}
is the  Lagrangian of the CC interaction of neutrinos, leptons and $W$-bosons. Taking into account arbitrariness of the phases of fermion fields, we can put $CP$ phase factors of the lepton and neutrino fields equal to one. We have
\begin{equation}\label{CPtrans}
V_{CP}~l_{L}(x)~V^{-1}_{CP}=\gamma^{0}~C~\bar l_{L}^{T}(x'),\quad
V_{CP}~\nu_{iL}(x)~V^{-1}_{CP}=\gamma^{0}~C~\bar\nu_{iL}^{T}(x').
\end{equation}
From these relation we find
\begin{equation}\label{CPtrans1}
V_{CP}~\bar l_{L}(x)~V^{-1}_{CP}=-l^{T}_{L}(x')~C^{-1}\gamma^{0},
\quad V_{CP}~\bar
\nu_{iL}(x)~V^{-1}_{CP}=-\nu^{T}_{iL}(x')~C^{-1}\gamma^{0}.
\end{equation}
For the field of the charged $W^{\pm}$ vector bosons we have
\begin{equation}\label{CPtrans2}
V_{CP}~W_{\alpha}(x)~V^{-1}_{CP}=-\delta_{\alpha}~W^{\dag}_{\alpha}(x'),
\end{equation}
where  $\delta_{\alpha}$ is a sign factor ($\delta_{0}=1, ~\delta_{i}=-1$).
From all these relations we easily find
\begin{eqnarray}\label{CPinv1}
V_{CP}~\mathcal{L}_{I}^{CC}(x)~V^{-1}_{CP}&=&
-\frac{g}{\sqrt{2}}\sum_{l,i}\bar
\nu_{iL}(x')~\gamma_{\alpha}~U_{li}~l_{L}(x')~W^{\alpha}(x')\nonumber\\
&-&\frac{g}{\sqrt{2}}\sum_{l,i} \bar
l_{L}(x')~\gamma_{\alpha}~U^{*}_{li}~\nu_{iL}(x')~W^{\alpha\dag}(x').
\end{eqnarray}
Comparing (\ref{CPinv}) and (\ref{CPinv1}) we come to the conclusion that in the case of the $CP$ invariance in the lepton sector the Dirac mixing matrix
is real
\begin{equation}\label{CPU}
U_{li}=U_{li}^{*}.
\end{equation}
We will consider now the case of the Majorana fields \cite{Wolf,BilNedPet,Kayser}. The $CP$ transformation of the Majorana field $\nu_{i}$ has the form
\begin{equation}\label{CPMjf}
V_{CP}~\nu_{i}(x)~V^{-1}_{CP}=\eta^{*}_{i}~\gamma^{0}~C~\bar
\nu_{i}^{T}(x')=\eta^{*}_{i}~\gamma^{0}~\nu_{i}(x'),
\end{equation}
where $\eta^{*}_{i}$ is a phase factor. Unlike the Dirac fields, it can not be included in the field.  We will show now that the phase factor $\eta_{i}$ can take the values $\pm i$. In fact, from  (\ref{CPMjf}) by the hermitian conjugation and multiplication
from the right by the matrix $\gamma^{0}$  we find
\begin{equation}\label{CPMjf1}
V_{CP}~\bar \nu_{i}(x)~V^{-1}_{CP}=
\eta_{i}~\bar\nu_{i}(x')~\gamma^{0}.
\end{equation}
From this relation we have
\begin{equation}\label{CPMjf2}
V_{CP}~C~\bar \nu^{T}_{i}(x)~V^{-1}_{CP}=
\eta_{i}~C~\gamma^{0T}~C^{-1}~C~\bar\nu^{T}_{i}(x')=-
\eta_{i}~\gamma^{0}~C~\bar\nu^{T}_{i}(x').
\end{equation}
Finally, taking into account the Majorana condition we find
\begin{equation}\label{CPMjf3}
V_{CP}~ \nu_{i}(x)~V^{-1}_{CP}= -\eta_{i}~\gamma^{0}~\nu_{i}(x').
\end{equation}
If we  compare now (\ref{CPMjf}) and  (\ref{CPMjf3}) we conclude that
\begin{equation}\label{MjCPpar}
\eta^{*}_{i}=-\eta_{i},\quad \eta^{2}_{i}=-1.
\end{equation}
Thus, the $CP$ parity of a Majorana field can take  values $\pm i$.

From (\ref{CPinv}), (\ref{CPtrans}) and (\ref{CPMjf}) we find that in the case of the $CP$ invariance in the lepton sector the Majorana mixing matrix satisfies the condition
\begin{equation}\label{CPMjcond}
U_{li}~\eta^{*}_{i}=U^{*}_{li}.
\end{equation}

Finally we will obtain the standard parametrization of the $3\times 3$ Dirac mixing matrix. Let us consider two systems of orthogonal and normalized vectors
$|i\rangle$ and $|\nu_{l}\rangle$ ($i=1,2,3,~~l=e,\mu,\tau$). We have
\begin{equation}\label{3vec}
\langle k|i\rangle=\delta_{ik}, \quad \langle l'|l\rangle =\delta_{l'l}.
\end{equation}
Vectors $|\nu_{l}\rangle$ and $|i\rangle$ are connected by the relation
\begin{equation}\label{3vec1}
|\nu_{l}\rangle=\sum_{i}U^{*}_{li}|i\rangle.
\end{equation}
From (\ref{3vec}) it is obvious that $U$ is an unitary matrix.

In the most general case vectors $|\nu_{l}\rangle$ can be obtained from vectors $|i\rangle$ by three Euler rotations. The first rotation will be performed at the
angle $\theta_{12}$ around the vector $|3\rangle $. New orthogonal
and normalized vectors are
\begin{eqnarray}
&|1\rangle^{(1)}&= c_{12}~|1\rangle
+s_{12}~|2\rangle\nonumber\\
&|2\rangle^{(1)}&= -s_{12}~|1\rangle +c_{12}~|2\rangle\nonumber\\
&|3\rangle^{(1)}&=|3\rangle, \label{1rot}
\end{eqnarray}
where $c_{12}=\cos\theta_{12}$ and $s_{12}=\sin\theta_{12}$.
In the
matrix form (\ref{1rot}) can be written as follows
\begin{equation}\label{1rot1}
|\nu\rangle^{(1)}=U^{(1)}~|\nu\rangle.
\end{equation}
Here
\begin{eqnarray}\label{1rotcol}
|\nu\rangle^{(1)}=\left(%
\begin{array}{c}
 |1\rangle^{(1)}\\
 |2\rangle^{(1)}\\
  |3\rangle^{(1)}\\
\end{array}%
\right),~~~~
|\nu\rangle=\left(%
\begin{array}{c}
 |1\rangle\\
 |2\rangle\\
  |3\rangle\\
\end{array}%
\right)
\end{eqnarray}
and
\begin{eqnarray}\label{U1rot}
U^{(1)}=\left(%
\begin{array}{ccc}
  c_{12} & s_{12}& 0\\
  -s_{12} & c_{12} & 0\\
  0 & 0 & 1\\
\end{array}%
\right)
\end{eqnarray}
We will perform now the  second rotation at the angle $\theta_{13}$
around the vector $|2 \rangle^{(1)}$. At this step we will introduce
the $CP$ phase $\delta$. We have
\begin{equation}\label{2rot1}
|\nu\rangle^{(2)}=U^{(2)}~|\nu\rangle^{(1)}.
\end{equation}
Here
\begin{eqnarray}\label{U2rot}
U^{(2)}=\left(%
\begin{array}{ccc}
  c_{13} & 0& s_{13}e^{i\delta}\\
  0 & 1 & 0\\
  -s_{13}e^{-i\delta} & 0 &c_{13}  \\
\end{array}
\right).
\end{eqnarray}
Finally, let us perform the rotation around the vector $|1\rangle^{(2)}$
at the angle $\theta_{23}$. We have
\begin{equation}\label{3rot1}
|\nu^{\rm{mix}}\rangle=U^{(3)}~|\nu\rangle^{(2)}.
\end{equation}
Here
\begin{eqnarray}\label{U3rot}
|\nu^{\rm{mix}}\rangle=\left(%
\begin{array}{c}
  |\nu_{e}\rangle\\
 | \nu_{\mu}\rangle  \\
 |\nu_{\tau}\rangle  \\
\end{array}%
\right)
\end{eqnarray}
and
\begin{eqnarray}\label{U3rot}
U^{(3)}=\left(%
\begin{array}{ccc}
  1 & 0& 0\\
  0 & c_{23}  & s_{23} \\
  0 & -s_{23}  &c_{23}  \\
\end{array}%
\right).
\end{eqnarray}
From (\ref{1rot1}), (\ref{2rot1}) and (\ref{3rot1}) we find
\begin{equation}\label{Urot}
    |\nu^{\rm{mix}}\rangle=U^{*}~|\nu\rangle,
\end{equation}
where
\begin{eqnarray}\label{unitmixU}
U=(U^{(3)}~U^{(2)}~U^{(1)})^{*}=
\left(%
\begin{array}{ccc}
  1 & 0& 0\\
  0 & c_{23}  & s_{23} \\
  0 & -s_{23}  &c_{23}  \\
\end{array}%
\right)
\left(%
\begin{array}{ccc}
  c_{13} & 0& s_{13}e^{-i\delta}\\
  0 & 1 & 0\\
  -s_{13}e^{i\delta} & 0 &c_{13}  \\
\end{array}%
\right)
\left(%
\begin{array}{ccc}
  c_{12} & s_{12}& 0\\
  -s_{12} & c_{12} & 0\\
  0 & 0 & 1\\
\end{array}%
\right)
\end{eqnarray}
This is so-called the standard parametrization of the  3$\times$3 Dirac mixing
matrix. This matrix is characterized by three mixing angles
$\theta_{12}$, $\theta_{23}$ and $\theta_{13}$ and the $CP$ phase $\delta$. From
(\ref{unitmixU}) we have
\begin{eqnarray}
U=\left(\begin{array}{ccc}c_{13}c_{12}&c_{13}s_{12}&s_{13}e^{-i\delta}\\
-c_{23}s_{12}-s_{23}c_{12}s_{13}e^{i\delta}&
c_{23}c_{12}-s_{23}s_{12}s_{13}e^{i\delta}&c_{13}s_{23}\\
s_{23}s_{12}-c_{23}c_{12}s_{13}e^{i\delta}&
-s_{23}c_{12}-c_{23}s_{12}s_{13}e^{i\delta}&c_{13}c_{23}.
\end{array}\right).
\label{unitmixU1}
\end{eqnarray}
The 3$\times$3 Majorana mixing matrix is characterized by three mixing angles and three $CP$ phases. It can be presented in the form
\begin{equation}\label{3Mjmix}
    U^{M}=U~S^{M}(\alpha),
\end{equation}
where the matrix $U$ is given by (\ref{unitmixU}) and
\begin{eqnarray}\label{3Mjcol}
 S^{M}(\alpha)=\left(
\begin{array}{c}
e^{i\alpha_{1}}\\
e^{i\alpha_{2}}\\
1\\
\end{array}
\right),
\end{eqnarray}
where $\alpha_{1,2}$ are additional Majorana phases.

\section{On neutrino oscillations}

The most important  manifestation of the neutrino mixing are
neutrino oscillations. Neutrino oscillations are based on the fact
that in processes of neutrino production and neutrino detection due
to Heisenberg uncertainty principle small neutrino mass-squared
differences can not be resolved. As a result, in a weak decay
\begin{equation}\label{decays}
a\to b +l^{+}+\nu_{l}
\end{equation}
together with the lepton $l^{+}$ a "mixed" left-handed flavor
neutrino $\nu_{l}$ is produced. The state of $\nu_{l}$ is {\em a
coherent superposition} of the states of neutrinos with definite
masses
\begin{equation}\label{flavorstate}
|\nu_{l}\rangle = \sum_{i} U_{l i}^* ~ |\nu_i\rangle,
\end{equation}
where $|\nu_i\rangle$ is the  state of neutrino with mass $m_{i}$ and
momentum $p_{i}$.

If at t=0  flavor neutrino $\nu_{l}$ is produced, at the time t for
the neutrino state  we have
\begin{equation}\label{3neutrinostate}
|\nu_{l}\rangle _{t} = e^{-i\,H\, t }\, |\nu_{l}\rangle =
\sum^{3}_{i=1}\,e^{-i\,E_{i}\, t }\,U^{*}_{li}~|\nu_{i} \rangle
=\sum_{l'}|\nu_{l'}\rangle \sum^{3}_{i=1} U_{l'i} \,e^{-i\,E_{i}\, t
}\,U^{*}_{li}.
\end{equation}
Thus, the probability of the transition $\nu_{l} \to \nu_{l'}$
during the time interval t is given by the expression
\begin{equation}\label{3probability}
P(\nu_{l} \to \nu_{l'})=| \sum^{3}_{i=1}U_{l'i}
\,e^{-i\,(E_{i}-E_{k})\, t }\,U^{*}_{li}|^{2},
\end{equation}
where $k$ is fixed. If all phase  differences are small
($|E_{i}-E_{k}|\, t\ll 1$) or/and there is no mixing
($U^{*}_{li}=\delta_{li}$) in this case  it will be no neutrino oscillations ($P(\nu_{l} \to
\nu_{l'})\simeq \delta_{l'l}$). Thus,
neutrino oscillations are effect of the neutrino mixing and relatively large phase difference(s).

Assuming that $\vec{p_{i}}=\vec{p}$,~ we obtain the standard
expression for the phase difference
\begin{equation}\label{phasedifference}
(E_{i}-E_{k})~t\simeq \frac{\Delta m^{2}_{ki}}{2E}~L.
\end{equation}
Here $\Delta m^{2}_{ki}=m^{2}_{i}-m^{2}_{k}$ and $L\simeq t$ is the
distance between a neutrino source and neutrino detector.

The transition probability $P(\nu_{l} \to \nu_{l'})$ depends
on six parameters (two mass-squared differences $\Delta m^{2}_{23}$
and  $\Delta m^{2}_{12}$, three mixing angles $\theta_{23}$,
$\theta_{12}$ and  $\theta_{13}$ and  CP phase $\delta$). However,
from analysis of the data of neutrino oscillation experiments
follows that the parameter $\sin^{2}\theta_{13}$ and the ratio
$\frac{\Delta m^{2}_{12}}{\Delta m^{2}_{23}}$ are small:
\begin{equation}\label{smallparam}
\frac{\Delta m^{2}_{12}}{\Delta m^{2}_{23}}\simeq 3\cdot
10^{-2},\quad \sin^{2}\theta_{13}\lesssim 5\cdot 10^{-2}.
\end{equation}
If we neglect contribution of  the small parameters to the transition
probabilities,  we will find that
in the atmospheric and accelerator long baseline  region of the values of the parameter $\frac{L}{E}$ the two-neutrino $\nu_{\mu}\rightleftarrows \nu_{\tau}$ oscillations, driven
by $\Delta m^{2}_{23}$, take place. From (\ref{3probability}) and (\ref{phasedifference}) for the
probability of $\nu_{\mu}$ to survive we obtain the following
expression (see, \cite{BGG})
\begin{equation}\label{atmsur}
{\mathrm P}(\nu_{\mu} \to \nu_{\mu})\simeq 1-\frac {1}
{2}~\sin^{2}2\theta_{23}~(1-\cos\Delta m^2_{23} \frac {L} {2E})~.
\end{equation}
In the KamLAND region   $\nu_{e}\rightleftarrows \nu_{\mu,\tau}$
oscillations, driven by $\Delta m^{2}_{12}$, take place in the
leading approximation. For the probability of $\bar\nu_{e}$ to
survive we obtain the following expression
(see \cite{BGG})
\begin{equation}\label{2nuKL}
{\mathrm P} (\bar\nu_{e} \to \bar\nu_{e}) \simeq 1-\frac {1}
{2}~\sin^{2}2\theta_{12}~(1-\cos\Delta m^2_{12} \frac {L} {2E})~.
\end{equation}
 In the leading approximation the probability of solar
$\nu_{e}$ to survive in matter is also  given by the two-neutrino expression. It depends on $\tan^{2}\theta_{12}$, $\Delta m^2_{12}$ and electron
number density in the sun.

We will present now the results of the analysis of the  experimental data. From the analysis of the data of the {\em atmospheric Super-Kamiokande
experiment} for the parameters $\Delta m^{2}_{23}$ and $\sin^{2}2
\theta_{23}$ the following 90 \% CL ranges were obtained \cite{SK}
\begin{equation}\label{SKrange}
 1.5\cdot 10^{-3}\leq \Delta m^{2}_{23} \leq 3.4\cdot
10^{-3}\rm{eV}^{2},\quad \sin^{2}2 \theta_{23}> 0.92.
\end{equation}
The results of the atmospheric Super-Kamiokande experiment were
confirmed by the  K2K \cite{K2K} and MINOS \cite{Minos}
accelerator long-baseline neutrino oscillations experiments. From
the analysis of the MINOS data for the neutrino oscillation
parameters the following values were found \cite{Minos}
\begin{equation}\label{Minosdata}
\Delta m^{2}_{23}=(2.43 \pm 0.13)\cdot
10^{-3}\rm{eV}^{2},\quad \sin^{2}2 \theta_{23}>0.90~(90\% ~CL).
\end{equation}
From the  global analysis of the data of the reactor KamLAND
experiment  and data of  the solar neutrino experiments for the
parameters $\Delta m^{2}_{12}$ and $\tan^{2} \theta_{12}$ the
following values were obtained \cite{Kamland}
\begin{equation}\label{KLsolar}
\Delta m^{2}_{12} = (7.59^{+0.21}_{-0.21})\cdot
10^{-5}~\rm{eV}^{2},\quad\tan^{2} \theta_{12}= 0.47^{+0.06}_{-0.05}
\end{equation}
In the reactor CHOOZ experiment \cite{Chooz} no indications in favor
of $\bar\nu_{e} \to \bar\nu_{e}$ transitions, driven by $\Delta
m^2_{23}$, were found. From the exclusion plot, obtained from the
data of this experiment,  for the parameter $\sin^{2}\theta_{13}$
the following upper bound can be inferred
\begin{equation}\label{teta13}
\sin^{2}\theta_{13}\lesssim  5\cdot 10^{-2}~.
\end{equation}
At present a stage of the high precision neutrino oscillation experiments starts. In the future DOUBLE CHOOZ \cite{Doublechooz}, Daya
Bay \cite{Dayabay} and RENO \cite{Reno} reactor neutrino experiments sensitivities to the
parameter $\sin^{2}2 \theta_{13}$ will be  (10-20) times better
than in the CHOOZ experiment. The same sensitivity is planned to be
reached in the accelerator T2K experiment \cite{T2K}. In this
experiment parameters $\Delta m^{2}_{23}$ and $ \sin^{2}2
\theta_{23}$ will be measured with the accuracies $\delta(\Delta
m^{2}_{23})\sim 10^{-4}\rm{eV}^{2}$ and $\delta (\sin^{2}2
\theta_{23})\sim 10^{-2}$, correspondingly. High precision neutrino
oscillation experiments are planned at the future Super Beam \cite{Superbeam},
Beta-beam \cite{Betabeam}, and Neutrino Factory facilities \cite{Nufac}.

\section{Basic elements of the  theory of $0\nu\beta\beta$-decay}

In this section we will consider   the neutrinoless double
$\beta$-decay of even-even nuclei \cite{DoiKatani,BilPet}
\begin{equation}\label{betabeta}
(A,Z) \to (A,Z+2) + e^{-} + e^{-}.
\end{equation}
We will assume that
\begin{itemize}
  \item The Hamiltonian of the weak interaction is given by the  SM.
  \item The neutrino mixing takes place.
  \item Neutrinos with definite masses {\em $\nu_{i}$ are Majorana particles}.
\end{itemize}
For the effective Hamiltonian of the process we have
\begin{equation}\label{effham}
{\mathcal{H}}_{I}(x)= \frac{G_F}{\sqrt{2}} 2~\sum_{i}\bar e_{L}(x)
\gamma_{\alpha}~ U_{li}~\nu_{iL}(x)~j^{\alpha}(x) + \mathrm{h.c.}
\end{equation}
Here $G_F$ is the Fermi constant, $j^{\alpha}(x)$ is the hadronic
charged current and the field $\nu_{i}(x)$
satisfies the condition
\begin{equation}\label{MJcond}
\nu^{c}_{i}(x)=C \bar\nu^{T}_{i}(x)=\nu_{i}(x).
\end{equation}
The neutrinoless double $\beta$-decay is the second order in
$G_{F}$ process with the virtual neutrinos.
The matrix element of the process  is given by the
following expression
\begin{eqnarray}\label{Smatelem}
&&\langle f|S^{2}|i
\rangle=4\frac{(-i)^{2}}{2~!}~\left (\frac{G_F}{\sqrt{2}}\right )^{2}N_{p_1}N_{p_2}
 \int
 \sum_{i}\bar
u_{L}(p_1)e^{ip_{1}x_{1}}\gamma_{\alpha}~U_{ei}\langle 0|T(\nu_{iL}(x_{1})~\nu^{T}_{iL}(x_{2})|0\rangle
\nonumber\\
&&\times \gamma^{T}_{\beta}~U_{ei}\bar
u^{T}_{L}(p_2)e^{ip_{2}x_{2}}\langle
N_{f}|T(J^{\alpha}(x_{1})J^{\beta}(x_{2}))|N_{i}
\rangle~d^{4}x_{1}d^{4}x_{2}-(p_{1}\rightleftarrows p_{2}).
\end{eqnarray}
Here $p_{1}$ and $p_{2}$ are electron momenta, $J^{\alpha}(x)$
is the weak charged current in the Heisenberg representation, $N_{i}$ and $N_{f}$ are the states of the initial and the final nuclei
with 4-momenta $P_{i}=(E_{i}, \vec{p_{i}})$ and  $P_{f}=(E_{f}, \vec{p_{f}})$, respectively, and $N_{p}=\frac{1}{(2\pi)^{3/2}\sqrt{2p^{0}}}$ is the standard
normalization factor.

Let us consider the neutrino propagator. From the Majorana condition (\ref{MJcond}) we find
\begin{equation}\label{nupropag1}
    \langle 0|T(\nu_{iL}(x_{1})
\nu^{T}_{iL}(x_{2})|0\rangle
=-\frac{1-\gamma_{5}}{2}\langle
0|T(\nu_{i}(x_{1})
\bar\nu_{i}(x_{2}))|0\rangle~\frac{1-\gamma_{5}}{2}~C.
\end{equation}
Further, we have
\begin{equation}\label{nupropag2}
 \langle
0|T(\nu_{i}(x_{1})
\bar\nu_{i}(x_{2}))|0\rangle=\frac{i}{(2\pi)^{4}}
\int e^{-iq~(x_{1}-x_{2})}\frac{\gamma\cdot q+m_{i}}{q^{2}-m^{2}_{i}} d^{4}q
\end{equation}
Thus, we for the neutrino propagator we find the following expression
\footnote{Notice that in the case of the Dirac
neutrinos $\langle 0|\nu_{iL}(x_{1})
\nu^{T}_{iL}(x_{2})|0\rangle=\frac{1-\gamma_{5}}{2}~\langle
0|\nu_{i}(x_{1})
\nu^{T}_{i}(x_{2})|0\rangle\frac{1-\gamma^{T}_{5}}{2}=0$. The
neutrinoless double $\beta$-decay is obviously forbidden in the Dirac
case.}
\begin{equation}\label{nupropag3}
 \langle
0|T(\nu_{iL}(x_{1})
\bar\nu_{iL}(x_{2}))|0\rangle=-\frac{i}{(2\pi)^{4}}
\int e^{-iq~(x_{1}-x_{2})}\frac{m_{i}}{q^{2}-m^{2}_{i}}~ d^{4}q~\frac{1-\gamma_{5}}{2}~C.
\end{equation}
The neutrino propagator is proportional to $m_{i}$. It is obvious from (\ref{nupropag2}) that this is connected with the fact that only left-handed
neutrino fields enter into the Hamiltonian of the weak interaction.
In the case of massless neutrinos ($m_{i}=0, ~i=1,2,3$), in accordance with the theorem on the equivalence of the theories with massless
Majorana and Dirac neutrinos, the  matrix element of the neutrinoless double $\beta$-decay is equal to zero.

Let us consider the second term of the matrix element (\ref{Smatelem}).
It is easy to show  that
\begin{eqnarray}\label{relation}
\bar u_{L}(p_1)\gamma_{\alpha} (1-\gamma_{5})\gamma_{\beta}C \bar
u^{T}_{L}(p_2)&=&\bar u_{L}(p_2)C^{T}\gamma^{T}_{\beta}
(1-\gamma^{T}_{5})\gamma^{T}_{\alpha} \bar
u^{T}_{L}(p_1)\nonumber\\&=& -\bar u_{L}(p_2)\gamma_{\beta}
(1-\gamma_{5})\gamma_{\alpha}C \bar u^{T}_{L}(p_1)~.
\end{eqnarray}
If we take into account (\ref{relation}) and the  relation
\begin{equation}\label{relation1}
T(J^{\beta}(x_{2})J^{\alpha}(x_{1}))=T(J^{\alpha}(x_{1})J^{\beta}(x_{2}))
\end{equation}
we can show that the second term of the matrix element (\ref{Smatelem}) is equal to the first one. Thus, for the matrix element
we obtain the following expression
\begin{eqnarray}\label{Smatelem1}
\langle f|S^{2}|i
\rangle&=&-4 \left (\frac{G_F}{\sqrt{2}}\right )^{2}N_{p_1}N_{p_2}
 \int\bar
u_{L}(p_1)e^{ip_{1}x_{1}}\gamma_{\alpha}\frac{i}{(2\pi)^{4}}
\sum_{i}U^{2}_{ei}m_{i}\int \frac{e^{-iq~(x_{1}-x_{2})}}
{ p^{2}-m^{2}_{i}}d^{4}q\nonumber\\
&&\times \frac{1-\gamma_{5}}{2}\gamma_{\beta}C~\bar
u^{T}_{L}(p_2)e^{ip_{2}x_{2}} \langle
N_{f}|T(J^{\alpha}(x_{1})J^{\beta}(x_{2}))|N_{i}
\rangle~d^{4}x_{1}d^{4}x_{2}
\end{eqnarray}
Initial  nuclei in the process (\ref{betabeta}) are
$^{76}\rm{Ge}$, $^{136}\rm{Xe}$, $^{130}\rm{Te}$, $^{100}\rm{Mo}$
and other heavy nuclei. The calculation of the nuclear part of  the matrix element of
the $0\nu\beta\beta$-decay is a complicated nuclear problem. In
such a calculation different approximations are used. We will present
now the matrix element of the $0\nu\beta\beta$-decay in a form which
is  appropriate  for such approximate calculations.

Let us perform in (\ref{Smatelem1}) the integration  over the
time variables $x^{0}_{2}$ and $x^{0}_{1}$. The integral over
$x^{0}_{2}$ can be presented in the form
\begin{equation}\label{integration}
    \int^{\infty}_{-\infty}
...dx^{0}_{2}=\int_{-\infty}^{x^{0}_{1}}...dx^{0}_{2}+\int_{x^{0}_{1}}^{\infty}...
    dx^{0}_{2}~.
\end{equation}
After the integration over  $q^{0}$ in the neutrino propagator, in the region  $x^{0}_{1}>x^{0}_{2}$ we find\footnote{It is assumed that in the propagator  $m^{2}_{i} = m^{2}_{i}-i\epsilon$.}
\begin{equation}\label{nupropag4}
\frac{i}{(2\pi)^{4}}~\int\frac{e^{-iq~(x_{1}-x_{2})}} {
q^{2}-m^{2}_{i}}d^{4}q=\frac{1}{(2\pi)^{3}}~
\int\frac{e^{-iq^{0}_{i}~(x^{0}_{1}-x^{0}_{2})+i\vec{q}~
(\vec{x}_{1}-\vec{x}_{2})}}{2~q_{i}^{0}}d^{3}q~,
\end{equation}
where
\begin{equation}\label{energy}
q_{i}^{0}=\sqrt{\vec{q}^{2}+m^{2}_{i}}~.
\end{equation}
In the region $x^{0}_{1}<x^{0}_{2}$ we have
\begin{equation}\label{nupropag5}
\frac{i}{(2\pi)^{4}}~\int\frac{e^{-iq~(x_{1}-x_{2})}} {
q^{2}-m^{2}_{i}}d^{4}q=\frac{1}{(2\pi)^{3}}~
\int\frac{e^{-iq^{0}_{i}~(x^{0}_{2}-x^{0}_{1})+i\vec{q}~
(\vec{x}_{2}-\vec{x}_{1})}}{2~q_{i}^{0}}d^{3}q.
\end{equation}
For the operators $J^{\alpha}(x)$ from the invariance under the translations  we have
\begin{equation}\label{Heisen1}
J^{\alpha}(x)=e^{iHx^{0}}J^{\alpha}(\vec{x})e^{-iHx^{0}},
\end{equation}
where $H$ is the total Hamiltonian. From this
relation we find
\begin{eqnarray}\label{phases}
&&\langle N_{f}|J^{\alpha}(x_{1})J^{\beta}(x_{2})|N_{i}
\rangle=\nonumber\\&&\sum_{n}
e^{i(E_{f}-E_{n})x^{0}_{1}}e^{i(E_{n}-E_{i})x^{0}_{2}} ~\langle
N_{f}|J^{\alpha}(\vec{x_{1}})|N_{n}\rangle\langle N_{n}|
J^{\beta}(\vec{x_{2}}))|N_{i}\rangle~,
\end{eqnarray}
where $|N_{n}\rangle$ is the vector of the state of the intermediate
nucleus with 4-momentum $P_{n}=(E_{n}, \vec{p_{n}})$. In (\ref{phases}) the sum
over the total system of the states $|N_{n}\rangle$ is assumed.

Taking into account that  at $\pm\infty$ the interaction is turned
off we have
\begin{equation}\label{phases1}
 \int_{-\infty}^{0}e^{i a x^{0}_{2}}~dx^{0}_{2}
 \to \int_{-\infty}^{0}e^{i (a-i\epsilon) x^{0}_{2}}~dx^{0}_{2}
=\lim_{\epsilon\to 0}\frac{-i}{a-i\epsilon}
\end{equation}
and
\begin{equation}\label{phases2}
 \int^{-\infty}_{0}e^{i a x^{0}_{2}}~dx^{0}_{2}
 \to \int^{\infty}_{0}e^{i (a+i\epsilon) x^{0}_{2}}~dx^{0}_{2}
=\lim_{\epsilon\to 0}\frac{i}{a+i\epsilon}.
\end{equation}
From (\ref{phases1}) and (\ref{phases2}) we find
\begin{eqnarray}\label{phases3}
&&\int^{\infty}_{-\infty} dx^{0}_{1}\int_{-\infty}^{x^{0}_{1}}dx^{0}_{2}\sum_{n}\langle
N_{f}|J^{\alpha}(\vec{x_{1}})|N_{n}\rangle \langle N_{n}| J^{\beta}(\vec{x_{2}})|N_{i} \rangle
e^{i(E_{f}-E_{n})x^{0}_{1}+ i(E_{n}-E_{i})x^{0}_{2}}e^{i(p^{0}_{1}x^{0}_{1}+p^{0}_{2}x^{0}_{2})}\times
\nonumber\\
&&e^{iq_{i}^{0}(x^{0}_{2}-x^{0}_{1})}=-i\sum_{n} \frac{\langle N_{f}|J^{\alpha}(\vec{x_{1}})|N_{n}\rangle\langle N_{n}|
J^{\beta}(\vec{x_{2}}))|N_{i}}{E_{n}+p^{0}_{2}+q^{0}_{i}-E_{i}-i\epsilon}
~2\pi\delta(E_{f}+p^{0}_{1}+p^{0}_{2}-E_{i})
\end{eqnarray}
Taking into account all these relations, for the matrix element of the neutrinoless
double $\beta$-decay we obtain the following expression
\begin{eqnarray}\label{Smatelem2}
&&\langle f|S^{2}|i
\rangle=2i~\left(\frac{G_F}{\sqrt{2}}\right )^{2}N_{p_1}N_{p_2}
 \bar
u(p_1)\gamma_{\alpha}\gamma_{\beta}(1+\gamma_{5})C \bar
u^{T}(p_2)\int
d^{3}x_{1}d^{3}x_{1}e^{-i\vec{p_{1}}\vec{x_{1}}-i\vec{p_{2}}\vec{x_{2}}}\times \nonumber\\
&&\sum_{i}U^{2}_{ei}m_{i}\frac{1}{(2\pi)^{3}}
\int\frac{e^{i\vec{q}~(\vec{x_{1}}-\vec{x_{2}})}} {
q_{i}^{0}}d^{3}q[\sum_{n} \frac{\langle
N_{f}|J^{\alpha}(\vec{x_{1}})|N_{n}\rangle\langle N_{n}|
J^{\beta}(\vec{x_{2}}))|N_{i}\rangle }{E_{n}+p^{0}_{2}+q^{0}_{i}-E_{i}-i\epsilon}
\nonumber\\ &&+\sum_{n}\frac{ \langle
N_{f}|J^{\beta}(\vec{x_{2}})|N_{n}\rangle\langle N_{n}|
J^{\alpha}(\vec{x_{1}}))|N_{i}\rangle}{E_{n}+p^{0}_{1}+q^{0}_{i}-E_{i}-i\epsilon}
]~2\pi\delta(E_{f}+p^{0}_{1}+p^{0}_{2}-E_{i})
\end{eqnarray}
The equation (\ref{Smatelem2}) is the exact expression for the matrix
element of $0\nu\beta\beta$-decay in the second order of the
perturbation theory. We will consider major $0^{+}\to 0^{+}$
transitions of even-even nuclei. For such transitions the following
approximations are standard.
\begin{enumerate}
  \item Small neutrino masses  can be safely
neglected in   $q^{0}_{i}$.

The averaged momentum of the virtual
neutrino is given by the relation $q\simeq
  \frac{1}{r}$, where $r$ is  the average distance between two nucleons in
  nucleus. Taking into account that $r \simeq 10^{-13}$ cm, we have  $q \simeq 100$ MeV. Neutrino masses are smaller than 2.2 eV. Thus, we have $q^{0}_{i}=\sqrt{\vec{q}^{2}+m^{2}_{i}}\simeq q$

\item Long-wave approximation.

We have $p_{k}x_{k}\leq p_{k}R$, where $R\simeq 1.2~
A^{1/3}\cdot10^{-13}$ cm is the radius of nucleus ($k=1,2$). Taking
into account that $p_{k}\lesssim 1$ Mev, we have $p_{k}x_{k}\ll 1$.
Thus, $e^{-i\vec{p_{1}}\vec{x_{1}}-i\vec{p_{2}}
\vec{x_{2}}}\simeq 1$ i.e. two electrons are emitted in $S$-states.

\item Closure approximation.

Energy of the virtual neutrino is much larger than
the excitation energy $(E_{n}-E_{i})$.
 Thus,  we can change the energy of the
intermediate states $E_{n}$ by average energy $\overline{ E}$.  In this
(closure) approximation we have
\begin{equation}\label{Smatelem3}
\frac{\langle
N_{f}|J^{\alpha}(\vec{x_{1}})|N_{n}\rangle\langle N_{n}|
J^{\beta}(\vec{x_{2}}))|N_{i}\rangle }{E_{n}+p^{0}_{2}+q^{0}_{i}-E_{i}-i\epsilon}\simeq \frac{\langle
N_{f}|J^{\alpha}(\vec{x_{1}})
J^{\beta}(\vec{x_{2}}))|N_{i}\rangle }{\overline{ E}+p^{0}_{2}+q-E_{i}-i\epsilon}.
\end{equation}

\item The impulse approximation for the hadronic charged
current $J^{\alpha}(\vec{x})$.

Taking
into account the major terms, the hadronic charged
current takes the form \footnote{The pseudoscalar term in
 the one-nucleon matrix element of the hadronic charged current induces a tensor term in the current. From numerical calculations follow that its contribution to the matrix element can be significant (see \cite{Simkovic99}).}
\begin{equation}\label{impCC}
J^{\alpha}(\vec{x})\simeq
\sum_{n}\delta(\vec{x}-\vec{r}_{n})~\tau^{n}_{+}[~g_{V}(q^{2})g^{\alpha
0}+g_{A}(q^{2})\sigma^{n} _{i}g^{\alpha i}]~.
\end{equation}
Here $g_{V}(q^{2})$ and $g_{A}(q^{2})$ are vector and axial
formfactors ,  $\sigma_{i}$ and $\tau_{i}$ are Pauli matrices,
$\tau_{+}=\frac{1}{2}~(\tau_{1}+i \tau_{2})$ and index $n$ runs over
all nucleons in a nucleus. We have $g_{V}(0)=1,~~g_{A}(0)=g_{A}\simeq
1.27$.
\end{enumerate}
It is obvious that $\tau^{n}_{+}~\tau^{n}_{+}=0$. Thus,   in the impulse approximation the hadronic currents satisfy the relation
\begin{equation}\label{impCC1}
J^{\alpha}(\vec{x}_{1})~J^{\beta}(\vec{x}_{2})=
J^{\beta}(\vec{x}_{2})~J^{\alpha}(\vec{x}_{1}).
\end{equation}
Further, the matrix
$\gamma_{\alpha}\gamma_{\beta}$ in the leptonic part of the matrix
element (\ref{Smatelem2}) can be presented in the form
\begin{equation}\label{gamrel}
\gamma_{\alpha}\gamma_{\beta}=g_{\alpha\beta}+\frac{1}{2}~
(\gamma_{\alpha}\gamma_{\beta}-
\gamma_{\beta}\gamma_{\alpha}).
\end{equation}
It follows from (\ref{impCC1}) that the second term of
(\ref{gamrel}) does not give contribution to the matrix element.
From (\ref{impCC}) we have
\begin{equation}\label{CCprod}
J^{\alpha}(\vec{x_{1}})J_{\alpha}(\vec{x_{2}})=\sum_{n,m}\tau^{n}_{+}\tau^{m}_{+}
\delta(\vec{x}_{1}-\vec{r}_{n})~\delta(\vec{x}_{2}-\vec{r}_{m}) (
g^{2}_{V}(q^{2})
-g^{2}_{A}(q^{2})~\vec{\sigma}^{n}\cdot\vec{\sigma}^{m})~,
\end{equation}
Neglecting nuclei recoil,   we obtain  in the laboratory frame
$$M_{i}=M_{f}+p^{0}_{2}+p^{0}_{1},$$
where $M_{i}$ and $M_{f}$ are masses of the initial and final nuclei. From this relation we find
\begin{equation}\label{edenom}
q+p^{0}_{1,2}+\overline{ E}-M_{i}=q\pm
(\frac{p^{0}_{1}-p^{0}_{2}}{2})+\overline{ E}-\frac{M_{i}+M_{f}}{2}
\end{equation}
The  term $(\frac{p^{0}_{1}-p^{0}_{2}}{2})$, is much smaller
than all other terms in the right-hand side of this relation. Neglecting this term, we have
\begin{equation}\label{edenom1}
q+p^{0}_{1,2}+\overline{ E}-M_{i}\simeq q
+\overline{ E}-\frac{M_{i}+M_{f}}{2}
\end{equation}
Further, taking into account that $g_{V}(q^{2})\simeq \frac{1}{1+ \frac{q^{2}}{0.71 ~ \mathrm{GeV}^{2}}}$ and $g_{A}(q^{2})\simeq \frac{1}{1+ \frac{q^{2}}{M^{2}_{A}}}$, where
 $M_{A}\simeq 1~ \mathrm{GeV}^{2}$,  we can neglect $q^{2}$-dependence of the formfactors. After the integration in the matrix element (\ref{Smatelem2}) over $\vec{x_{1}}$ and  $\vec{x_{2}}$,
 for the neutrino propagator we find the following expression
\begin{equation}\label{1neuprop}
    \frac{1}{(2\pi)^{3}}
\int\frac{e^{i\vec{q}~\vec{r}_{nm}}d^{3}q} {q(
q+ \overline{ E}-\frac{1}{2}(M_{i}+M_{f}))}=\frac{1}{4\pi R}H(r_{nm},\overline{ E}),
\end{equation}
where
\begin{equation}\label{2neuprop}
H(r,\overline{ E})=\frac{2R}{\pi r}~\int^{\infty}_{0} \frac{\sin qr~dq}{q+\overline{ E}-\frac{1}{2}(M_{i}+M_{f})}.
\end{equation}
Here $R$ is the nuclei radius and $\vec{r}_{nm}=\vec{r}_{n}-\vec{r}_{m}$.

Taking into account all these relations, from ({\ref{Smatelem2}) for the matrix element of $0\nu\beta\beta$-decay we obtain the following expression
\begin{eqnarray}\label{Smat6}
\langle f|S^{2}|i
\rangle&=&-i~\left(\frac{G_F}{\sqrt{2}}\right )^{2}\frac{1}{(2\pi)^{3}}~\frac{1}
{\sqrt{p^{0}_{1}p^{0}_{2}}}
~m_{\beta\beta}~g^{2}_{A}~\frac{1}{R}~ \bar u(p_1)
(1+\gamma_{5})C\bar u^{T}(p_2)\nonumber\\&&\times M^{0\nu}~~
\delta(p^{0}_{1}+p^{0}_{2}+M_{f}-M_{i}),
\end{eqnarray}
where
\begin{equation}\label{efMjm}
m_{\beta\beta}=\sum_{i}U^{2}_{ei}m_{i}
\end{equation}
is the effective Majorana mass and
\begin{equation}\label{NME}
M^{0\nu}=M^{0\nu}_{GT}-\frac{1}{g^{2}_{A}}~M^{0\nu}_{F}
\end{equation}
is the nuclear matrix element. Here
\begin{equation}\label{Fme}
M^{0\nu}_{F}= \langle
\Psi_{f}|\sum_{n,m}H(r_{n,m},\overline{E})~\tau^{n}_{+}\tau^{m}_{+}|\Psi_{i}
\rangle
\end{equation}
is the Fermi matrix element and
\begin{equation}\label{GTme}
M^{0\nu}_{GT}= \langle
\Psi_{f}|\sum_{n,m}H(r_{n,m},\overline{E})~
\tau^{n}_{+}\tau^{m}_{+}~\vec{\sigma}^{n}\cdot\vec{\sigma}^{m})|\Psi_{i}
\rangle
\end{equation}
is the Gamov-Teller matrix element. In (\ref{Fme}) and (\ref{GTme})
$|\Psi_{i,f}
\rangle$ are wave function of the initial and final nuclei.

From (\ref{Smat6})  we conclude that
 matrix element of $0\nu\beta\beta$-decay is a
product of the effective Majorana mass $m_{\beta\beta}$, the electron
matrix element and the nuclear matrix element which includes neutrino
propagator (neutrino potential). Taking into account that $\overline{ E}-\frac{1}{2}(M_{i}+M_{f}$ is much smaller than $\bar q$ for the neutrino propagator we obtain the following approximate relation
\begin{equation}\label{hfunc2}
H(r)\simeq \frac{2R}{\pi}\int^{\infty}_{0} \frac{\sin
qr}{qr}dq=\frac{R}{r}~.
\end{equation}
 Using the standard rules, from (\ref{Smat6}) we can easily obtain the decay rate of the
$0\nu\beta\beta$-decay. The electron part of the decay probability is given by the trace
\begin{equation}\label{Tr}
\rm{Tr} (1+\gamma_{5})(\gamma\cdot
p_{2}-m_{e})(1-\gamma_{5})(\gamma\cdot p_{1}+m_{e})=8p_{1}p_{1}~.
\end{equation}
Taking into account the final state electromagnetic interaction of the
electrons and nucleus for the decay rate of the
$0\nu\beta\beta$-decay we find the following expression
\begin{eqnarray}\label{decrate}
d\Gamma^{0\nu}&=&
|m_{\beta\beta}|^{2}~|M^{0\nu}|^{2}~\frac{1}{(2\pi)^{5}}~G_{F}^{4}~\frac{1}{R^{2}}
~g^{4}_{A}(E_{1}E_{2}-p_{1}p_{2}\cos\theta)\times
\nonumber\\&&~F(E_{1},(Z+2))~F(E_{2},(Z+2))~p_{1}p_{2}~ \sin\theta
d\theta~dE_{2},
\end{eqnarray}
where $E_{1,2}\equiv p^{0}_{1,2}$ is electron total energy
($E_{2}=M_{i}-M_{f}-E_{1})$, $\theta$ is the angle between electron
momenta $\vec {p}_{1}$ and $\vec {p}_{2}$ and
\begin{equation}\label{Ffunc}
F(Z)\simeq \frac{2\pi\eta}{1-e^{-2\pi\eta}}~,
\end{equation}
is the Fermi function ($\eta=Z\alpha~\frac{m_{e}}{p}$).

From (\ref{decrate}) follows that for the ultra relativistic electrons
$\theta$-dependence of the decay rate is given by the factor $(1-\cos\theta)$.
Thus, ultra relativistic electrons can not be emitted in the same
direction. This is connected with the fact that the  helicity of the high
energy electrons, produced in the weak interaction, is
 equal to -1 . If
electrons are emitted in the same direction, the projection of their
total angular momentum on the direction of the momentum is equal to -1. It is obvious
that such electrons can not be produced in $O^{+}\to O^{+}$
transition.

From expression (\ref{decrate}) for the total decay rate we obtain
the following expression
\begin{equation}\label{totrate}
\Gamma^{0\nu}=\frac{1}{T^{0\nu}_{1/2}}=|m_{\beta\beta}|^{2}~|M^{0\nu}|^{2}~
G^{0\nu}(Q,Z),
\end{equation}
where\footnote{An additional factor $\frac{1}{2}$ is due to the fact that in the final state we have two identical electrons.}
\begin{eqnarray}\label{Gfac}
G^{0\nu}(Q,Z)&=&\frac{1}{2(2\pi)^{5}}~G_{F}^{4}~\frac{1}{R^{2}}
~g^{4}_{A}\int_{0}^{Q}dT_{1}~\int^{\pi}_{0}\sin\theta
d\theta~(E_{1}E_{2}-p_{1}p_{2}\cos\theta)p_{1}p_{2}\times\nonumber\\&&
F(E_{1},(Z+2))~F(E_{2},(Z+2)).
\end{eqnarray}
Here $T_{1}=E_{1}-m_{e}$,  $Q=M_{i}-M_{f}-2m_{e}$ is the
total released kinetic energy and $T^{0\nu}_{1/2}$ is the half-life
of the $0\nu\beta\beta$-decay. In the Table I we present numerical
values of $G^{0\nu}(Q,Z)$ for some nuclei \cite{Rodin03}.
\begin{center}
 Table I
\end{center}
\begin{center}
The values of the factor $G^{0\nu}(Q,Z)$ for some nuclei
\end{center}
\begin{center}
\begin{tabular}{|c|c|}
  \hline  Nucleus &  $G^{0\nu}(Q,Z)$ in units $10^{-25}y^{-1}\rm{eV}^{-2}$\\
\hline   $^{76}\rm{Ge}$& 0.30
\\
\hline   $^{100}\rm{Mo}$ & 2.19
\\
\hline   $^{130}\rm{Te}$ & 2.12
\\
\hline   $^{136}\rm{Xe}$ & 2.26
\\
\hline
\end{tabular}
\end{center}
The total rate of the $0\nu\beta\beta$-decay is the product of
three factors:
\begin{enumerate}
  \item The modulus squared of the effective Majorana mass.
   \item Square of nuclear matrix element.
  \item The known factor $G^{0\nu}(Q,Z)$.
\end{enumerate}
We have considered in some details neutrinoless double $\beta$-decay
of nuclei
\begin{equation}\label{2betadecay}
(A,Z) \to (A,Z+2)+e^{-} +e^{-}.
\end{equation}
There could be other second order in the Fermi constant $G_{F}$ processes with the virtual Majorana neutrinos in which the total lepton number is changed by two. The examples are the decays
\begin{equation}\label{Kdecay}
 K^{-}\to \pi^{+}+\mu^{-} +e ^{-}
\end{equation}
and
\begin{equation}\label{Kdecay1}
  K^{+}\to \pi^{-}+\mu^{+} +\mu ^{+}
\end{equation}
the process
\begin{equation}\label{mue}
\mu^{-}+(A,Z) \to (A,Z-2)+e^{+}
\end{equation}
and others.

The leptonic part of the operator which give contribution to matrix
elements of  (\ref{2betadecay}),  (\ref{Kdecay})
and other similar processes is given by
\begin{equation}\label{operator}
\sum_{i}T(\bar
l_{L}(x_{1})\,\gamma_{\alpha}~
U_{li}~\langle 0|~T(\nu_{iL}(x_{1})\,\nu^{T}_{iL}(x_{2}))\, |0 \rangle\,U_{l'i}
\,\gamma^{T}_{\beta} \,\bar l'^{T}_{L}(x_{2})),\quad l,l'=e,\mu,
\end{equation}
where Majorana neutrino propagator is given by the expression (\ref{nupropag3}). Taking into account that $m^{2}_{i} \ll p^{2}$, we can neglect $m^{2}_{i}$ in the denominator of the propagator. Thus, the matrix element of a process in which a lepton pair $(ll')$ is produced, is proportional to
\begin{equation}\label{effMjll'}
 m_{ll'} = \sum_{i}U_{li}\,U_{l'i}\,m_{i}.
\end{equation}
Analogously, matrix elements of the processes (\ref{Kdecay1}), (\ref{mue}) and other similar processes are proportional to
$ m^{*}_{ll'}$.

The sensitivities to the parameter  $| m_{ll'}|$ of the experiments on the search for the processes (\ref{Kdecay}), (\ref{Kdecay1}), (\ref{mue})
 and other similar processes  are much worse than the sensitivity of the experiments on the search for $0\nu\beta\beta$-decay to the parameter
$| m_{\beta\beta}|$.

For example, in the experiment \cite{Kaulard} on the search for the process
$\mu^{-}\rm{Ti}\to e^{+}\rm{Ca}$ the following upper bound was obtained
\begin{equation}\label{upper}
\frac{\Gamma(\mu^{-}\rm{Ti}\to e^{+}\rm{Ca})}{\Gamma(\mu^{-}\rm{Ti}\to \rm{all})}
\leq 1.7 \cdot10^{-12}.
\end{equation}
For the probability of the decay
$K^{+}\to \pi^{-}\mu^{+} \mu^{+}$ the following upper bound was reached \cite{Appel}:
\begin{equation}\label{upper1}
\frac{\Gamma(K^{+}\to \pi^{-}\mu^{+} \mu^{+})}{\Gamma(K^{+}\to \rm{all})}\leq 3\cdot 10^{-9}.
\end{equation}
From these data the following upper bounds can be found (see \cite{ElliotVogel})
\begin{equation}\label{bound2}
|m_{\mu e}|\leq 82~ \rm{MeV};~~|m_{\mu \mu}|\leq 4\cdot 10^{4}~ \rm{MeV}.
\end{equation}
These values must be compared with the sensitivity of the  experiments on the search for $0\nu\beta\beta$-decay
to the effective Majorana mass (in today's experiments $| m_{\beta\beta}|\simeq (0.2-1.3)~\mathrm{eV}$ (see below)).

The effective Majorana mass is determined by neutrino masses and neutrino mixing angles. An information about the neutrino mixing angles $\theta_{ik}$ and
neutrino mass-squared differences $\Delta m^{2}_{ik}$ was obtained from the data of
the neutrino oscillation experiments. Taking into account these data, we
will consider now  possible values of the effective Majorana mass.

\section{Effective Majorana mass}
From neutrino oscillation data  follows that one mass-squared
difference (solar) is much smaller than the other one
(atmospheric). For three massive neutrinos two types
of neutrino mass spectra are possible in this case.
\begin{enumerate}
\item Normal spectrum
\begin{equation}\label{norspec}
m_{1}<  m_{2}    <  m_{3} ;\quad \Delta m^{2}_{12}  \ll    \Delta
m^{2}_{23}
\end{equation}
\item Inverted spectrum\footnote{In order to have the same notation
$\Delta m^{2}_{12}$ for
the solar-KamLAND
neutrino mass-squared difference and to determine this quantity as a positive one the neutrino masses are usually labeled differently in the cases of
the normal and inverted neutrino mass spectra.
 In the case of the normal
spectrum $\Delta m^{2}_{23}>0$ and in the case of  the inverted
spectrum $\Delta m^{2}_{13}<0$. Thus, with such a notation  the character of the neutrino mass spectrum is
determined by the sign of the larger (atmospheric)  neutrino
mass-squared difference. It is clear, however, that the sign  of the
atmospheric  mass-squared difference has no physical meaning: it
is a convention based on the labeling of the neutrino masses and
determination of the neutrino mass-squared difference ($\Delta m^{2}_{ik}= m^{2}_{k}- m^{2}_{i}$). In both
cases of the neutrino mass spectrum for the mixing angles the same notations
can be used.}
\begin{equation}\label{invspec}
m_{3}<  m_{1} <  m_{2} ;\quad \Delta m^{2}_{12}  \ll
 | \Delta m^{2}_{13}|
\end{equation}
\end{enumerate}
In the case of the normal spectrum the neutrino masses $m_{2,3}$ are
connected with the lightest mass $m_{1}$ and two neutrino mass-squared
differences $\Delta m^{2}_{12}$ and $\Delta m^{2}_{23}$ by the
following relations
\begin{equation}\label{norspec1}
m_{2}=\sqrt{m^{2}_{1}+\Delta m^{2}_{12}},~~
m_{3}=\sqrt{m^{2}_{1}+\Delta m^{2}_{12}+\Delta m^{2}_{23}}
\end{equation}
In the case of the inverted spectrum we have
\begin{equation}\label{invspec1}
m_{1}=\sqrt{m^{2}_{3}+|\Delta m^{2}_{13}|},\quad
m_{2}=\sqrt{m^{2}_{3}+|\Delta m^{2}_{13}|+\Delta m^{2}_{12}}
\end{equation}
It is obvious that effective Majorana mass is determined not only by the lightest neutrino mass and neutrino mass-squared differences but also by the character of the neutrino mass spectrum.

Usually the following three typical neutrino mass spectra  are considered\footnote{Let us notice that these three neutrino mass spectra correspond to
different mechanisms of neutrino mass generation. Masses of quarks
and charged leptons satisfy hierarchy of the type (\ref{hierar}).
Hierarchy of neutrino masses is a typical feature of GUT models
(like $SO(10)$) in which quarks and leptons are unified. Inverted
spectrum and quasi-degenerate spectrum require specific symmetries
of the neutrino mass matrix.}
\newpage
\begin{enumerate}
  \item  Hierarchy of the neutrino masses
\begin{equation}\label{hierar}
m_{1} \ll m_{2} \ll m_{3}.
\end{equation}
\item  Inverted hierarchy of the neutrino masses
  \begin{equation}\label{invierar}
m_{3} \ll m_{1} <m_{2}
\end{equation}
 \item
Quasi-degenerate neutrino mass spectrum
\begin{equation}\label{quasi}
 m_{1}\simeq
m_{2}\simeq m_{3},\quad m_{1}(m_{3})\gg\sqrt{ \Delta m^{2}_{\rm{23}}}~(\sqrt{ |\Delta
m^{2}_{\rm{13}}|}).
\end{equation}

\end{enumerate}
We will discuss now the possible values of the effective Majorana mass in the case of these three neutrino mass spectra.
\begin{center}
I.  Hierarchy of the neutrino masses
\end{center}
In this case
we have
\begin{equation}\label{hierar1}
m_{1} \ll \sqrt{\Delta m^{2}_{12}},\quad m_{2}\simeq \sqrt{ \Delta
m^{2}_{12}},\quad m_{3}\simeq  \sqrt{ \Delta m^{2}_{23}}.
\end{equation}
Thus, in the case of neutrino mass hierarchy the neutrino masses
$m_{2}$ and $m_{3}$ are determined by the neutrino mass-squared
differences $\Delta m^{2}_{12}$ and $\Delta m^{2}_{23}$, correspondingly, and the lightest mass is very small. Neglecting the
contribution of $m_{1}$ to   the effective Majorana mass and using the standard parametrization of the neutrino mixing matrix
we find
\begin{equation}\label{hierar2}
|m_{\beta\beta}|\simeq\left |\,
 \sin^{2} \theta_{12}\, \sqrt{\Delta m^{2}_{12}} + e^{2i\,\alpha}
 \sin^{2} \theta_{13}\, \sqrt{\Delta m^{2}_{23}}\,\right |~.
\end{equation}
Here $\alpha$ is a Majorana phase difference.

The first term in Eq.(\ref{hierar2}) is small because of the
smallness of $\Delta m^{2}_{12}$. The contribution of the ``large''
$\Delta m^{2}_{23}$ to $|m_{\beta\beta}|$ is suppressed by
the small factor $\sin^{2} \theta_{13} $. Using the values (\ref{Minosdata})
and (\ref{KLsolar}) and the CHOOZ bound (\ref{teta13}), we have
\begin{equation}\label{hierar3}
    \sin^{2} \theta_{12}\, \sqrt{\Delta
m^{2}_{12}}\simeq 2.8\cdot 10^{-3}~\mathrm{eV},\quad
\sin^{2} \theta_{13}\,
\sqrt{\Delta m^{2}_{23}}\lesssim  2.5\cdot 10^{-3}\rm{eV}.
\end{equation}
Thus, if the value of the parameter $\sin^{2}\theta_{13}$ is close to the CHOOZ bound, the
first term and the modulus of the second term of (\ref{hierar2}) are
approximately equal and at $\alpha\simeq \pi/2$
 the terms in the expression (\ref{hierar2})
practically  cancel each other. In this case the Majorana mass
$|m_{\beta\beta}|$ will be close to zero.

Even without this possible
cancelation  the effective
Majorana mass in the case of the neutrino mass hierarchy is very small. In fact, from (\ref{hierar2}) and (\ref{hierar3})
 we have the following upper bound
\begin{equation}\label{hierar4}
|m_{\beta\beta}|\leq \left (\sin^{2} \theta_{12}\, \sqrt{\Delta
m^{2}_{12}}+\sin^{2} \theta_{13}\, \sqrt{\Delta m^{2}_{23}}\right )\lesssim  5.3
\cdot 10^{-3}~\rm{eV}.
\end{equation}
This bound  is significantly smaller that the expected
sensitivity of the future experiments on the search for
$0\nu\beta\beta$-decay (see later).

\begin{center}
 II. Inverted hierarchy of the neutrino masses
\end{center}

For the neutrino masses we have in this case
\begin{equation}\label{invierar1}
m_{3}\ll \sqrt{ |\Delta m^{2}_{\rm{13}}|},~~m_{1}\simeq \sqrt{
|\Delta m^{2}_{13}|},~ m_{2}\simeq\sqrt{ |\Delta m^{2}_{13}|}~(1+
\frac{\Delta m^{2}_{12}}{2\, |\Delta m^{2}_{13}|}).
\end{equation}
In the expression for the effective Majorana mass $|m_{\beta\beta}|$
the lightest mass $m_{3}$ is multiplied by the small parameter
$\sin^{2}\theta_{13}$. Neglecting the contribution of this term and
also neglecting the small term $\frac{\Delta m^{2}_{12}}{2\, |\Delta m^{2}_{13}|}$ in (\ref{invierar1})
 we find
\begin{equation}\label{invierar2}
|m_{\beta\beta}|\simeq \sqrt{| \Delta m^{2}_{13}|}\,~ (1-\sin^{2}
2\,\theta_{12}\,\sin^{2}\alpha)^{\frac{1}{2}},
\end{equation}
where $\alpha$ is the difference of the
Majorana phases of the elements $U_{e2}$ and $U_{e1}$. The phase
difference $\alpha$ is the only unknown parameter in the expression
for $|m_{\beta\beta}|$ in the case of the inverted hierarchy.
 From (\ref{invierar2}) we find
\begin{equation}\label{invierar3}
\cos  2\,\theta_{12} \,\sqrt{ |\Delta m^{2}_{13}|} \leq
|m_{\beta\beta}| \leq\sqrt{ |\Delta m^{2}_{13}|}.
\end{equation}
The upper and lower bounds of the inequality (\ref{invierar3}) corresponds to the $CP$-invariance in the lepton sector. In fact, the elements of the first row of the neutrino mixing matrix can be written in the form $U_{ei}=|U_{ei}|~e^{i\alpha_{i}}$. In the case of the $CP$-invariance, the elements of the neutrino mixing matrix satisfies the condition (\ref{CPMjcond}). From this condition we have
\begin{equation}\label{CPMaj}
e^{2i\alpha_{i}}=\eta_{i},
\end{equation}
where $\eta_{i}=\pm i$ is the $CP$ parity of the Majorana neutrino with mass $m_{i}$. For the phase difference $\alpha=\alpha_{2}-\alpha_{1}$
we have
\begin{equation}\label{CPMaj1}
e^{2i\alpha}=\eta_{2}~\eta^{*}_{1}.
\end{equation}
If $\eta_{2}=\eta_{1}$ we obtain $\alpha=0,\pi$ (the upper bound in the inequality (\ref{invierar3})). If  $\eta_{2}=-\eta_{1}$ we have $\alpha=\pm \frac{\pi}{2}$ (the lower bound in the inequality (\ref{invierar3})).

From (\ref{Minosdata}) and (\ref{KLsolar})  we find the following
 range of the possible values of the effective Majorana mass
\begin{equation}\label{invierar5}
1.8\cdot 10^{-2}\leq |m_{\beta\beta}|\leq 4.9\cdot 10^{-2}~\rm{eV}
\end{equation}
Thus, in the case of the inverted hierarchy of
the neutrino masses the lower bound of the effective Majorana mass is
different from zero.

The anticipated sensitivities to the effective Majorana mass of the next
generation of the experiments on the search for the $0\nu\beta\beta$-decay are
in the range (\ref{invierar5}) (see below). Thus, the future
$0\nu\beta\beta$-decay experiments will probe the Majorana nature
of neutrinos with definite masses in the case of the inverted hierarchy of the neutrino masses.
\begin{center}
III. Quasi-degenerate neutrino mass spectrum
\end{center}

Neglecting the small contribution of $\sin^{2}\theta_{13}$, for the
effective Majorana mass we obtain  in the case of the quasi-degenerate neutrino mass spectrum the following
expression
\begin{equation}\label{quasi1}
|m_{\beta\beta}|\simeq m_{\mathrm{min}}\, (1-\sin^{2}
2\,\theta_{\rm{12}}\,\sin^{2}\alpha)^{\frac{1}{2}},
\end{equation}
where $m_{\mathrm{min}}$ is  the lightest neutrino mass and $\alpha$ is
the Majorana phase difference. Thus, $|m_{\beta\beta}|$ depends in this case on two
unknown parameters: $m_{_{min}}$ and
$\alpha$. From (\ref{quasi1}) we  obtain the following range for
the effective Majorana mass:
\begin{equation}\label{quasi2}
\cos  2\,\theta_{\rm{12}} \,m_{\mathrm{min}} \leq |m_{\beta\beta}| \leq m_{\mathrm{min}}.
\end{equation}
If $0\nu\beta\beta$-decay will be observed and  the effective Majorana mass turn out to be relatively large ($|m_{\beta\beta}|\gg \sqrt{ \Delta m^{2}_{23}|}$)
it would be an evidence that neutrinos are Majorana particles and the spectrum of their mass is quasi-degenerate. In this case we could conclude that
\begin{equation}\label{quasi5}
|m_{\beta\beta}|\leq m_{\mathrm{min}}\leq  2.8\, |m_{\beta\beta}|.
\end{equation}

An information about the  lightest neutrino mass can be obtained from
experiments on the measurement of the end-point part of the  $\beta$-spectrum of  tritium. From existing data of the Mainz \cite{Mainz} and Troitsk \cite{Troitsk} tritium experiments it was found the upper bound
\begin{equation}\label{quasi3}
m_{\mathrm{min}}< 2.2~ \rm{eV}.
\end{equation}
The sensitivity of the future KATRIN experiment \cite{Katrin} is
expected to be
\begin{equation}\label{quasi4}
m_{\mathrm{min}}\simeq 0.2~\rm{ eV}
\end{equation}
We have considered three neutrino mass spectra with  special values of the lightest neutrino mass  $m_{\mathrm{min}}$. In Fig. 1 the effective Majorana mass for the normal and inverted neutrino mass spectra
as a function of $m_{\mathrm{min}}$
is presented. Uncertainties of the parameters $\Delta m^{2}_{12}$, $\Delta m^{2}_{23}$ and $\tan^{2}\theta_{12}$ and  possible values of the Majorana phase difference $\alpha$ are taken into account in Fig.1.


\begin{figure}[htp]
\centering
\includegraphics[width=8cm]{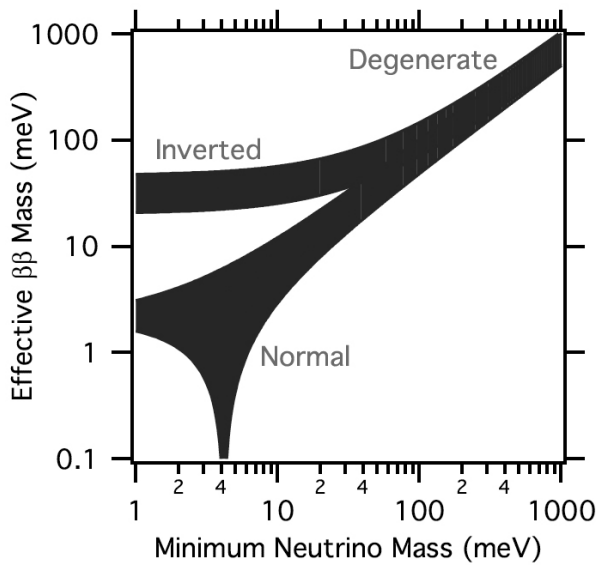}\\
\caption{The effective Majorana mass for the normal and inverted neutrino mass spectra as a function of minimal neutrino mass.}\label{fig:Fig 1}
\end{figure}

In conclusion let us notice  that if in the KATRIN (or other) experiments the neutrino mass will be measured and in the $0\nu\beta\beta$-decay experiments, sensitive to the effective Majorana mass in the range (\ref{quasi2}), a positive signal will not be observed it would be an evidence that neutrinos with definite masses are Dirac particles.

\section{Nuclear matrix elements of $0\nu\beta\beta$-decay }
The effective Majorana mass $|m_{\beta\beta}|$  is not directly
measurable quantity. From the measurement of the half-life of the
$0\nu\beta\beta$-decay only {\em the product of the effective
Majorana mass and the nuclear matrix element} can be obtained . In order to determine the effective Majorana mass we need to know nuclear matrix elements of the $0\nu\beta\beta$-decay (NME).

The calculation of NME is a complicated nuclear many body problem.
Two main approaches are used:
Nuclear Shell Model (NSM)\cite{NSM} and Quasiparticle Random Phase
Approximation (QRPA)\cite{QRPA1,QRPA2}.

The Nuclear Shell Model is attractive from physical point of view: there are many
spectroscopic data in favor of shell structure
of nuclei (spins and parities of nuclei,
binding energies of magic nuclei, etc.). It is based on the assumption that exist spherically symmetrical averaged nucleon field (usually oscillator potential) and one-particle states in this field are used as a basis for the description of valence nucleons.
An effective interaction between nucleons is taken into account in the Hamiltonian. Because of  computational difficulties, rather limited  number of one-particle
states can be used in the  NSM calculations. However, all possible distributions of valence nucleons over these states are taken into account.

The neutrinoless double $\beta$-decay of a nucleus is due to transition of two neutrons into two protons with the emission of two electrons. The operator of the transition of two neutrons into two protons can be presented in the form
of the sum of products of a operator of the absorption of two neutrons in a state with total momentum $J$ and parity $\pi$  and an operator of creation of two protons with the same momentum and parity :
\begin{equation}\label{fact}
 M=\sum (P^{J^{\pi}})^{\dag}P^{J^{\pi}}
\end{equation}
It was found \cite{NSM1} that the dominant contribution to the NME
comes from the  $0^{+}$ state of the neutron-neutron pair. Sizable contribution gives  also $2^{+}$ state. It is, however,  smaller and has opposite sign. The contributions of other states are negligibly small.  The dominance of the contribution of the  $0^{+}$ state corresponds to the pairing content of the initial and final wave functions. Let us notice that if seniority of the
initial and final wave functions is equal to zero NME would be maximal.

Further, it was found \cite{NSM2} that the major contribution to NME comes from pairs of  neutrons at the distance  $r\leq (2-3)$ fm. In order to take into account strong repulsion of nucleons at small distances ($\leq 1 $~fm) additional $r$-dependence (so called short-range correlations) is  introduced in the expression for the NME.

This additional $r$-dependence is parameterized by Jastrow-type function \cite{Jastrow}
\begin{equation}\label{Jastrow}
f(r)=1-e^{-ar^{2}}(1-br^{2}), \quad a=1.1 ~\mathrm{fm}^{-2},~b=0.68~\mathrm{fm}^{-2}.
\end{equation}
Recently it was proposed to take into account the short-range correlations
by an Unitary Correlation Operator Method (UCOM) \cite{UCOM}.
In this method the correlated wave function is obtained by an unitary transformation of uncorrelated wave function.

In the Table \ref{tab:1} we present the NME values of nuclear matrix elements of the $0\nu\beta\beta$-decay  $M^{0\nu}$ which were calculated with Jastrow-like and UCOM short-range correlations.
\begin{center}
 Table I
\end{center}
\begin{center}
The NSM values of nuclear matrix elements of the $0\nu\beta\beta$-decay  \cite{NSM2}\label{tab:1}
\end{center}
\begin{center}
\begin{tabular}{|c|c|c|}
  \hline  Nuclei transition &  $M^{0\nu}$(UCOM)&$M^{0\nu}$(Jastrow)\\
\hline   ${^{48}\rm{Ca}}\to{^{48}\rm{Ti}} $& 0.85&0.64
\\
\hline   ${^{76}\rm{Ge}}\to{^{76}\rm{Se}} $& 2.81& 2.30
\\
\hline   ${^{82}\rm{Se}}\to{^{82}\rm{Kr} }$& 2.64& 2.18
\\
\hline ${^{124}\rm{Sn}}\to{^{124}\rm{Te} }$& 2.62& 2.10
\\
\hline ${^{128}\rm{Te}}\to{^{128}\rm{Te} }$& 2.88& 2.34
\\
\hline ${^{130}\rm{Te}}\to{^{130}\rm{Xe} }$& 2.65& 2.12
\\
\hline ${^{136}\rm{Xe}}\to{^{136}\rm{Ba} }$& 2.19& 1.76
\\
\hline
\end{tabular}
\end{center}

Notice  that except double-magic nucleus ${^{48}\rm{Ca}}$  NSM nuclear matrix elements of the $0\nu\beta\beta$-decay  for all considered nuclei are practically the same (they differ by  not more than $\sim 20$ \%).

There are two groups which are performing QRPA calculation of NME at present: Tubingen group \cite{TUB1,TUB2,TUB3} and Jyvaskyla group \cite{Jyv,Jyv1,Jyv2,Jyv3}. QRPA method allows to include pairing correlations in nuclear wave functions trough the introduction of quasiparticles (particle-hole pairs). Two parameters $g_{pp}$ and
$g_{ph}$ of the model characterize particle-particle and particle-hole interactions. The constant $g_{ph}$ is obtained from the fit of the energy of the giant Gamov-Teller resonance. The Tubingen group determines the value of the constant $g_{pp}$
from the measured half-life of the $2\nu \beta\beta $-decay of the corresponding nucleus. The  Jyvaskyla    group
determines the constant $g_{pp}$
from data on the $\beta$-decay of
nuclei which are close to the nuclei
of the interest for the $0 \nu\beta \beta$-decay. They also use the value of
the constant $g_{pp}$, obtained from the half-life of the $2\nu \beta\beta $-decay.

In QRPA approach the mean nuclear field is described by the Woods-Saxon potential. The number of basic one-particle states which can be used  in the QRPA is much larger than in the NSM.
This is an important advantage of the QRPA approach. However, only limited excitations can be taken into account.

Like in the NSM case,  in the QRPA approach the  dominant contribution to NME  gives  $0^{+}$ state of neutron pairs.
However, in QRPA not only $2^{+}$ state  but also other states give significant contribution.

In both approaches main contribution to NME gives neutron pairs at the distance smaller than $(2-3)$ fm. The short-range correlations, taking into account nucleon repulsion at short distances, are introduced in the QRPA expression for NME via the  Jastrow-type function (\ref{Jastrow})
and through Unitary Correlation Operator Method procedure. Recently  \cite{TUB3} the short-range correlations were calculated directly from different nucleon-nucleon potentials by the coupled cluster method (CCM)\cite{CCM}.

In the Table II the results of the calculations of the QRPA nuclear matrix elements by the Tubingen  group are presented.  The
short-range correlations were calculated by CCM method.. For comparison in the Table II the results of the calculation of NME with the Jastrow-type short-range correlations are also presented.
\newpage

\begin{center}
 Table II
\end{center}
\begin{center}
The values of QRPA nuclear matrix elements of the $0\nu\beta\beta$-decay
with CCM and Jastrow short-range correlations \cite{TUB3}.
\end{center}
\begin{center}
\begin{tabular}{|c|c|c|}
  \hline  Nucleus&  $M^{0\nu}$ $(\mathrm{Jastrow})$&$M^{0\nu}(CCM)$\\
\hline   ${^{76}\rm{Ge}} $& 3.33~-~4.68& 4.07~-~6.64
\\
\hline   ${^{82}\rm{Se}}$& 2.82~-~4.17&3.53~-~5.92
\\
\hline ${^{96}\rm{Zr}}$&1.01~-~1.34&1.43~-~2.12
\\
\hline ${^{100}\rm{Mo}}$&2.22~-~3.53&2.94~-~5.56
\\
\hline ${^{100}\rm{Mo}}$&2.22~-~3.53&2.94~-~5.56
\\
\hline ${^{116}\rm{Cd}}$&1.83~-~2.93&2.30~-~4.14
\\
\hline ${^{128}\rm{Te}}$&2.46~-~3.77&3.21~-~5.65
\\
\hline ${^{130}\rm{Te}}$&2.27~-~3.38&2.92~-~5.04
\\
\hline ${^{136}\rm{Xe}}$&1.17~-~2.22&1.57~-~3.24
\\
\hline
\end{tabular}
\end{center}
The uncertainties of NME in Table II are mainly due to different values of the axial constant $g_{A}$ which are used in the calculations. Upper bounds of NME correspond to the free nucleon value
$g_{A}=1.25$ and lower bounds correspond to quenched in the nuclear matter value $g_{A}=1$.

The results of the calculations of nuclear matrix elements of the $0\nu\beta\beta$-decay
performed by the  Jyvaskyla    group are presented in the Table III.
The short-range correlations were taken into account by Jastrow and UCOM procedure.
\newpage
\begin{center}
 Table III
\end{center}
\begin{center}
The values of NME calculated in the framework of QRPA by the  Jyvaskyla    group  \cite{Jyv}
\end{center}
\begin{center}
\begin{tabular}{|c|c|c|c|c|}
  \hline  Nucleus & $g_{pp}$& $g_{A}$&$M^{0\nu}$$(\mathrm{Jastrow})$&$M^{0\nu}(UCOM)$\\
\hline   ${^{76}\rm{Ge}} $&1.02& 1.00& 5.08& 6.56
\\
&1.06& 1.25&  4.03& 5.36\\
\hline   ${^{82}\rm{Se}} $&0.96& 1.00& 3.54& 4.60
\\
&1.00& 1.25& 2.78& 3.72\\
\hline   ${^{96}\rm{Zr}} $&1.06& 1.00& 3.13& 4.31
\\
&1.11& 1.25& 2.07& 3.12\\
\hline   ${^{100}\rm{Mo}} $&1.07& 1.00& 3.53& 4.85
\\
&1.00& 1.25& 2.74& 3.93\\
\hline   ${^{116}\rm{Cd}} $&0.82($\beta$)& 1.25& 3.98& 4.93
\\
&0.97& 1.00& 3.68& 4.68\\
&1.01& 1.25& 3.03& 3.94\\
\hline   ${^{128}\rm{Te}} $&0.86($\beta$)& 1.25& 4.07& 5.51
\\
&0.89& 1.00& 4.23& 5.84\\
&0.92& 1.25& 3.38& 4.79\\
\hline   ${^{130}\rm{Te}} $&0.84& 1.00& 4.06& 5.44
\\
&0.90& 1.25& 2.99& 4.22\\
\hline   ${^{136}\rm{Xe}} $&0.74& 1.00& 2.86& 3.72
\\
&0.83& 1.25& 2.05& 2.80\\
\hline
\end{tabular}
\end{center}
It is difficult to expect that outcome of the many-body nuclear calculations, based on different assumptions, will be  the same. However, from the results presented in Tables I-III  we can conclude the following
\begin{enumerate}
  \item The values of the nuclear matrix elements of  the $0\nu\beta\beta$-decay of different nuclei obtained in the latest QRPA and NSM calculations are qualitatively  compatible.
  \item NSM nuclear matrix elements of ${^{76}\mathrm{Ge}}$, ${^{82}\mathrm{Se}}$ and ${^{130}\mathrm{Te}}$ are by a factor (1.5-2) lower
than  QRPA nuclear matrix elements.
\item There is no doubts that traditional methods of the calculation of NME  will be improved and, apparently, new methods will appear. However, it will be very important to find a  way to test the calculations.

If neutrinoless double $\beta$-decay will be discovered and half-live of {\em different nuclei}
is measured,
from the ratios of measured half-lives in this case  it will be possible to test different models of the calculation of NME \cite{NME}. If, for example, half-lives of the $0\nu\beta\beta$-decay of ${^{76}\mathrm{Ge}}$ and ${^{130}\mathrm{Te}}$
will be measured, the ratio of half-lives will be practically equal to the inverse ratio
of the corresponding phase-space factors in the case of NSM nuclear matrix elements and could  be significantly different from this ratio in the case of QRPA nuclear matrix elements.
\end{enumerate}

\section{Experiments on the search for  $0\nu\beta\beta$-decay}
At present exist data of many experiments on the search for neutrinoless double $\beta$-decay. The most stringent lower bound on the half-lives of
the $0\nu\beta\beta$-decay of different nuclei was obtained in the  Heidelberg-Moscow \cite{HMoscow} and  IGEX \cite{IGEX} experiments, and in the recent CUORICINO \cite{Cuoricino}
and   NEMO \cite{Nemo} experiments.

In the Heidelberg-Moscow and IGEX experiments two electrons with total energy $Q_{\beta\beta}=2039$ keV which are produced in the $0^{+}\to 0^{+}$  transition ${^{76}\mathrm{Ge}}\to
{^{76}\mathrm{Se}}+e^{-}+e^{-}$ were searched for. In the Heidelberg-Moscow experiment the source (and detector) consist of five crystals of 86 \% enriched ${^{76}\mathrm{Ge}}$ with total mass 10.96 kg. In the IGEX experiment $\sim$ 7 kg of enriched ${^{76}\mathrm{Ge}}$ was used. Low background level ($\sim$ 0.06 counts/(keV~kg~y)) and high energy resolution ( $\sim$ 3 keV) were reached in the germanium experiments.

For the half-life of ${^{76}\mathrm{Ge}}$ in the Heidelberg-Moscow experiment the following lower bound was obtained \cite{HMoscow}
\begin{equation}\label{HeiMos}
    T^{0\nu}_{1/2}({^{76}\mathrm{Ge}}> 1.9\cdot 10^{25}\mathrm{y}
\end{equation}
From this result
the following upper bound on the effective Majorana mass was inferred: $ |m_{\beta\beta}|<0.35$ eV.

 In the IGEX experiment it was found \cite{IGEX}
\begin{equation}\label{IGEX}
T^{0\nu}_{1/2}({^{76}\mathrm{Ge}}>1.57 \cdot 10^{25} \mathrm{y}.
\end{equation}
From this results, assuming different NME, it was found the bound: $ |m_{\beta\beta}|<(0.33-1.35)$ eV

In the cryogenic  experiment CUORICINO \cite{Cuoricino}
the search for the $0\nu\beta\beta$-decay of ${^{130}\mathrm{Te}}$
was performed. An array of 62 $\mathrm{TeO}_{2}$ crystals with a total active mass of 40.7 kg was cooled to
 $(8-10)$ mK in a dilution refrigerator. Since the heat capacity is proportional to $T^{3}$ an increase of temperature due to tiny release of
energy in the $0\nu\beta\beta$-decay can be recorded by special thermometers.

No evidence for the $0\nu\beta\beta$-decay of ${^{130}\mathrm{Te}}$ was obtained in the CUORICINO experiment. For the half-life of ${^{130}\mathrm{Te}}$ a limit
\begin{equation}\label{Cuoricino}
T^{0\nu}_{1/2}({^{130}\mathrm{Te}}> 3.0 \cdot 10^{24} \mathrm{y}.
\end{equation}
was obtained \cite{Cuoricino}. From this limit using the values of the NME, calculated in the latest papers, the following upper bound was inferred $|m_{\beta\beta}|<(0.19-0.68)$ eV.

In the NEMO3 experiment \cite{Nemo} the cylindrical source was divided in sectors with enriched ${^{100}\mathrm{Mo}}$ (6914 g), ${^{82}\mathrm{Se}}$ (932 g), ${^{116}\mathrm{Cd}}$ (405 g), ${^{130}\mathrm{Te}}$ (454 g),
${^{150}\mathrm{Nd}}$ (34 g), ${^{96}\mathrm{Zr}}$ (94 g) and ${^{48}\mathrm{Ca}}$ (7g g). For the detecting of the two electrons drift cells and plastic scintillator were used. No $0\nu\beta\beta$-decay was observed. In the Table IV the results of the NEMO3 experiment are presented.
\begin{center}
 Table IV
\end{center}
\begin{center}
Lower bounds of the half-lives of the $0\nu\beta\beta$-decay of different nuclei, obtained in the NEMO3 experiment\cite{Nemo}.
\end{center}
\begin{center}
\begin{tabular}{|c|c|c|}
  \hline  Nucleus& $ T^{0\nu}_{1/2}$ (90\%CL)&$|m_{\beta\beta}|$(eV)\\
\hline   ${^{100}\rm{Mo}} $& $\geqslant 5.8\cdot 10^{23}~\mathrm{y}$& $\leqslant(0.6-1.3)$
\\
\hline   ${^{82}\rm{Se}}$&$\geqslant 2.1\cdot 10^{23}~\mathrm{y}$&$\leqslant(1.2-2.2)$
\\
\hline ${^{96}\rm{Zr}}$&$\geqslant 8.6\cdot 10^{21}~\mathrm{y}$&$\leqslant(7.4-20.1)$
\\
\hline ${^{48}\rm{Ca}}$&$\geqslant 1.3\cdot 10^{22}~\mathrm{y}$&$\leqslant(29.7$
\\
\hline ${^{150}\rm{Nd}}$&$\geqslant 1.8\cdot 10^{22}~\mathrm{y}$&$\leqslant(4.0-6.3)$
\\
\hline
\end{tabular}
\end{center}
Several new experiments on the search for the $0\nu\beta\beta$-decay
are  at preparation at present. In these new experiments it is planned to reach the sensitivity  $|m_{\beta\beta}|\simeq \mathrm{a~ few}~10^{-2}$ eV , corresponding to the inverted hierarchy of the neutrino mass spectrum.

In the future GERDA experiment \cite{Gerda} array of enriched $\rm{Ge}$ crystals  will be cooled and shielded by liquid argon (or nitrogen) of very high radiopurity. In the phase I of the GERDA experiment 5 detectors from the Heidelberg-Moscow experiment (active mass 11.9 kg) and 3 detectors from the IGEX experiment (active mass 6 kg) will be used. The expected background at this phase of the experiment will be $\sim 10^{-2}$ counts/(kg~keV~y). The expected sensitivity will be $T_{1/2}({^{76}\mathrm{Ge}})\simeq 3\cdot 10^{25}$ y at 90\% CL.
Nonobservation of the neutrinoless double $\beta$-decay at this phase of the experiment would allow to obtain the upper bound $|m_{\beta\beta}|\lesssim 0.27$ eV (with QRPA NME).

During the phase II of the GERDA experiment additional 22 kg of the enriched $\mathrm{Ge}$ will be used (total active mass of the enriched $\mathrm{Ge}$  will be about 40 kg). The expected background $10^{-3}$ counts/(kg~keV~y).
The sensitivity $T_{1/2}({^{76}\mathrm{Ge}})\simeq 1.4\cdot 10^{26}$ y (at 90\% CL) is planned to be reached. This sensitivity corresponds to the sensitivity to the effective Majorana mass  $|m_{\beta\beta}|\simeq 0.11$ eV (QRPA NME).

If goals of  Phase I and Phase II will be achieved and the level of the background $10^{-4}$ counts/(kg~keV~y) will be reached it is planned (in collaboration with the Majorana Collaboration) to build $\sim$ 1 ton germanium detector with the aim to investigate
the region of the inverted neutrino mass hierarchy.

As it is well known, the group of participants of the Heidelberg-Moscow experiment claimed  that it found an evidence for neutrinoless double $\beta$-decay of
${^{76}\mathrm{Ge}}$~\cite{Klop}. For the half-life of the decay the authors obtained the following 3$\sigma$ range $T_{1/2}({^{76}\mathrm{Ge}})= (1.30-3.55)\cdot 10^{25}$ y. These values  correspond to the following
range for the effective Majorana mass
$|m_{\beta\beta}|=(0.24-0.58)$ eV. (with NME calculated in \cite{Muto}).
There is no detailed analysis of the systematic errors in
\cite{Klop} (see \cite{PDG}). The only way to confirm or refute the claim
is to perform more sensitive than the Heidelberg-Moscow experiment (preferably $^{76}\mathrm{Ge}$ experiment in order to avoid the NME problem).
One of the aim of the GERDA experiment is to check the claim made in
\cite{Klop}.

In the proposed Majorana experiment \cite{Majorana} an array of enriched $\mathrm{Ge}$ crystals
will be installed inside of high purity electroformed copper cryostat.
It is expected that the background in the Majorana experiment will be  a factor of 150 lower than in the Heidelberg-Moscow and IGEX experiments. Staged approach based of the 60 kg enriched $\mathrm{Ge}$ array (60/120/180 kg) is planned. The expected sensitivity at the first stage of the experiment ($T_{1/2}({^{76}\mathrm{Ge}})\simeq 5.5\cdot 10^{26}$ y ) will allow to check the claim made in the papers \cite{Klop}

In the cryogenic CUORE experiment \cite{Cuore} an array of 19 towers made from 5$\times$5$\times$5 ~$\mathrm{cm}^{3}$ $\mathrm{TeO_{2}}$ crystals is used as a source (detector).
The total number of the crystals in the experiment is equal to 988. The total mass of the crystals 741 kg of $\mathrm{TeO_{2}}$ (204 kg of $^{130}\mathrm{Te}$). In the CUORICINO experiment one similar tower of a mass 40.7 kg was used.

The expected background in the CUORE experiment is 0.01 counts/{kg ~keV~y). The expected sensitivity to the half-life is $T_{1/2}({^{130}\mathrm{Te}})\simeq 2.5\cdot 10^{26}$ y. With the present-day values of NME the following sensitivity to the effective Majorana mass will be achieved:
$|m_{\beta\beta}|\simeq (4.7-5.3)\cdot 10^{-2}$ eV.

In the future EXO experiment \cite{EXO} the $0\nu\beta\beta$-decay of ${^{136}\mathrm{Xe}}$ will be search for. Because there is no need to grow crystals and procedure of enrichment is relatively simple, $\mathrm{Xe}$ is ideal for a large scale (one ton or more) neutrinoless double $\beta$-decay experiment.
Ion ${^{136}\mathrm{Ba}^{++}}$, produced in the decay ${^{136}\mathrm{Xe}}\to {^{136}\mathrm{Ba}^{++}}+e^{-}+^e{-}$, by the capture of  an electron can be transferred to the ion ${^{136}\mathrm{Ba}^{+}}$ which is stable in ${\mathrm{Xe}}$. The EXO Collaboration plan to identify ${^{136}\mathrm{Ba}^{+}}$ ion by optical pumping with lasers. Single ion can be detected by this technique (via photon rate $10^{7}$/s}). When the program of the ${^{136}\mathrm{Ba}^{+}}$ tagging will be realized, the background in the experiment on the search for the $0\nu\beta\beta$-decay will be drastically reduced.

At present the EXO collaboration is constructing 200 kg liquid xenon TPC  with
${\mathrm{Xe}}$ enriched to 80\% in ${^{136}\mathrm{Xe}}$. No
${^{136}\mathrm{Ba}^{+}}$ tagging will be done at this stage. In this experiment the sensitivity $|m_{\beta\beta}|\simeq 1.5 \cdot 10^{-1}$ eV is anticipated.

We have discussed experiments on the search for neutrinoless double $\beta$-decay which will be done in the coming years. There are several other experiments which are in R \& D stage: Super-NEMO (${^{150}\mathrm{Nd}}$  or ${^{82}\mathrm{Se}}$)~\cite{SNemo}, MOON (${^{100}\mathrm{Mo}}$) \cite{Moon}, SNO++ (${^{150}\mathrm{Nd}}$)\cite{Sno+}, COBRA (${^{116}\mathrm{Cd}}$, ${^{130}\mathrm{Te}}$ )\cite{Cobra}, CANDLES (${^{48}\mathrm{Nd}}$)\cite{Candles}, DCBA (${^{150}\mathrm{Nd}}$)\cite{Dcba}, CAMEO (${^{116}\mathrm{Cd}}$)\cite{Cameo}, XMASS (${^{136}\mathrm{Xe}}$)\cite{Xmass} and others.

\section{Conclusion}
The observation of the neutrino oscillations in experiments with atmospheric, solar, reactor and accelerator neutrinos proves
that neutrino masses are different from zero and that the states of flavor neutrinos $\nu_{e}, \nu_{\mu}, \nu_{tau}$ are mixtures of states of neutrinos with different masses. There are two general possibilities for neutrinos with definite masses: they can be 4-component Dirac particles, possessing  conserved total lepton number which distinguish neutrinos and antineutrinos or purely neutral 2-component Majorana particles with identical neutrinos and antineutrinos.

It will be extremely important for the further development of the theory of the neutrino masses and mixing to answer the fundamental question: are neutrinos with definite masses Dirac or Majorana particles?

Neutrino masses are many orders of magnitude smaller than masses of their family partners, leptons and quarks. This fact tell us that {\em neutrino masses and masses of leptons and quarks have different origin.} The most natural possibility of the explanation of the smallness of the neutrino masses gives us the seesaw mechanism of the neutrino mass generation. This beyond the Standard Model mechanism connects smallness of neutrino masses with the violation of the total lepton number at a large scale and Majorana nature of neutrino masses. If it will be established that neutrinos with definite masses are Majorana particles it will be strong argument in favor of the seesaw origin of neutrino masses.

Investigation  of the neutrinoless double $\beta$-decay of nuclei is  the only practical way which could allow to proof that neutrinos are Majorana particles. This is simply connected with the fact that there are huge number of parent nuclei in a source. However, even if neutrinos are Majorana particle probabilities of the $0\nu\beta\beta$-decay is extremely small. There are two reasons for that
\begin{itemize}
  \item The $0\nu\beta\beta$-decay is the second order in the Fermi constant process.
  \item The $0\nu\beta\beta$-decay is possible due to neutrino helicity-flip. In the case of neutrino mixing this means that the matrix element of the process is proportional to effective Majorana mass $m_{\beta\beta}=\sum_{i}U^{2}_{ei}m_{i}$. Smallness of neutrino masses is additional suppression factor in the decay probability.
\end{itemize}
Experiments on the measurement of the half-lives of such rare process as neutrinoless double $\beta$-decay with severe requirements to background and energy resolution are extremely difficult. A big progress was achieved.
However, future experiments with about one ton detectors, which will allow to reach the region of values of the effective Majorana mass, which is predicted from neutrino oscillation data in the case of the inverted mass
hierarchy, is definitely a challenge. Taking into account importance of the problem of the nature of massive neutrinos, there is no doubts that goals of future experiments will be achieved.

In this review we considered $0\nu\beta\beta$-decay, driven by the left-handed SM weak interaction
and Majorana neutrino masses. If total lepton number is not conserved and neutrinos with definite masses are Majorana particles such mechanism of the $0\nu\beta\beta$-decay obviously must exist.
In the literature many other possible mechanisms of the $0\nu\beta\beta$-decay
were considered (for references see, for example, \cite{Avignone}). We shortly discuss here a mechanism due to the exchange of a heavy SUSY neutralino.
Let us assume that exist a R-parity and lepton number violating interaction which induce the transition $d\to u + \tilde{e}$ ($\tilde{e}$ is the selectron) . In combination with the standard SUSY interaction which induce transition $\tilde{e}\to e +\chi$ ($\chi$ is the neutralino) these two interactions in the case of the virtual neutralino provide the $0\nu\beta\beta$ transition $n+n\to p+p +e+e$. If the constants of the SUSY interactions are of the order of the electroweak constant $g$ and if masses of SUSY particles are characterized by a scale $\Lambda$ in this case a contribution of these interactions to the matric element of  the $0\nu\beta\beta$-decay is proportional to
\begin{equation}\label{susi}
    M_{SUSY}\sim G^{4}_{F}~\frac{m^{4}_{W}}{\Lambda^{5}}.
\end{equation}
This contribution must be compared with the contribution to the matrix element of the $0\nu\beta\beta$-decay of the standard small Majorana neutrino mass mechanism
\begin{equation}\label{standmech}
M_{0}\sim G^{4}_{F}~\frac{|m_{\beta\beta}|}{<q^{2}>}
\end{equation}
Taking into account that $|m_{\beta\beta}|\lesssim 1~\mathrm{eV}$ and
$<q^{2}>\simeq 100 ~\mathrm{MeV}^{2}$ we come to the conclusion that for $\Lambda\simeq 1$ TeV $M_{SUSY}$ can be comparable with $M_{0}$ if a hypothetical  SUSY interaction which does not conserve $R$-parity and the lepton number is characterized by the electroweak constant $g$ (for more details see \cite{Susy}).

I am grateful to Theory Department of TRIUMF for the hospitality and to S. Bacca and A. Schwenk for useful discussion of the problem of  nuclear matrix element of the $0\nu\beta\beta$-decay.

\appendix
\section {Ettore Majorana}
Great Italian physicist Ettore Majorana was born in Catania (Scicily, Italy) on 5.08.1906. His father was an engineer, specialist in telecommunication. There were five children in the family.\footnote{For a detailed biography of
E. Majorana see E. Amaldi \cite{Amaldi}}

In 1921 the family moved to Rome. In 1923 E. Majorana finished High School
and entered the Engineer Faculty of the Rome University.

Among  his fellow-students and friends were E. Segre. and E. Amaldi. In 1927 Segre and later Amaldi  transferred to Physics Faculty  and started to work
with E. Fermi who was appointed in 1926 as a Professor of theoretical physics
at Rome University.

E. Majorana was famous at Engineer faculty for his
extraordinary ability of solving difficult mathematical problems.
E. Segre convinced E. Majorana to meet and to speak with Fermi.
At that time Fermi was developing  the statistical model which is
known as Thomas-Fermi model. He explained Majorana the model and
showed him the table with numerical values of the screening
potential which he calculated numerically.

Next morning Majorana returned back to the Institute of Physics with his own table of values of the potential. He transformed second order nonlinear Thomas-Fermi equation into Riccati equation and solved it numerically. Majorana and Fermi results
coincided.

A few days later E.Majorana became student of the Physics Faculty. He
impressed everybody by his lively mind and broad interests. He was
very critical person. For his criticism he was called in the Fermi
group "Great Inquisitor".

In 1929 Majorana received diploma. His thesis were devoted to
the investigation of the structure of nuclei and to the theory of
the alpha-decay. His supervisor was Fermi.

After doctorate Majorana visited the Institute of Physics for a few hours every
day.  He  spend most of his time in library working and studying
Dirac, Heisenberg, Pauli, Weil and Wigner papers.

At that time Fermi and his group worked on problems of  atomic
and molecular physics. Majorana wrote six papers on the subject.
These papers demonstrated profound Majorana's ability of using
symmetry properties of the states. This allowed him to simplify the problem and to choose the suitable
approximation (which is normal now but was not usual at that time).
These papers also demonstrated perfect Majorana's knowledge  of experimental
data.

In 1932 Majorana received teaching diploma ("libero docente"). Committee (Fermi, Lo Surdo, Persico) concluded that "the candidate has
a complete mastery in theoretical physics".

In the end of 1931-beginning of 1932 Fermi and his group
started to concentrate their efforts on nuclear physics.

After discovery of the neutron by Chadwick (1932)
Majorana was one of the first who came to an idea that  constituents of nuclei are protons and neutrons. He started to develop the theory of nuclear forces. Majorana proposed the theory of space
exchange forces between p and n (Majorana potential).

Fermi was very interested in the idea and tried to convince Majorana to
publish his results. However, Majorana refused and even did not allow
Fermi to mention them in his talk at a conference
in Paris. E. Fermi  managed, however, to persuade Majorana to go to Leipzig
where  W. Heisenberg was working and to Copenhagen where N. Bohr.
was working.

E. Majorana was abroad during seven months, starting from January 1933.
Heisenberg, who worked at that time on the theory of nuclear forces, discussed with Majorana his paper on nuclear theory. He convinced Majorana to publish it.

After returning from Germany E. Majorana started to come to the
Institute of Physics at via Panisperna rather rare and after some months did
not come at all.

He was at home and became interested in
political economy, philosophy, construction of ships, medicine. He
even wrote a paper on statistical laws in physics and social
sciences which was discovered and published after his disappearance.

Meanwhile new talented physicists grown up in Italy (Wick, Racah, Giovanni Gentili Jr. and others). It was time to create a
new chair in theoretical physics. This chair was created at
the University of Palermo and in the beginning of 1937 a competition for the chair was announced.

It was a problem to convince Majorana to take part in the
competition. Finally, Fermi, Amaldi and  Segre   managed to convince him.

Majorana had no publications during several years. He sent to "Nuovo Cimento" his most important paper "Symmetrical
theory of the electron and the positron" in which the theory of the Majorana particles was proposed.

After that the following happened. By the request of Senator Giovanni
Gentili E. Majorana for his extraordinary abilities without competition was appointed as a professor at  Napoli University.

In  January 1938 E. Majorana came  to Napoli. In Napoli he had
rather lonely life. He went to the University
only when he had lectures (on Quantum mechanics).  After lectures he
visited Professor Carrelli with whom he
became friendly and discuss different problems in physics.
He  never mentioned what he was doing. He discussed his  neutrino theory
and Carrelli had an impression that Majorana considered this  theory as his most important contribution to physics.

On March 23 1938 E. Majorana decided to go to Palermo.
On March 25  Carrelli received a telegram from Majorana from Palermo.
He asked Carrelli  do not  worry about a letter which he would  receive.
In the letter which came soon, Majorana wrote that he found his life
useless and decided to commit suicide. Carrelli called  Fermi and Fermi called to Luciano, Ettore brother. Luciano   immediately went to Napoli. He understood that on
the evening of March 25 Ettore   took boat to Napoli. He was seen
sleeping in his cabin when the boat was entering into the  Napoli bay. He did not arrive to Napoli. His body was never found.

During several  months there was an investigation conducted by family and
by the police.  Vatican tried to find out whether he entered some monastery.
No traces were found.

I will finish with two citations:
"There are various kind of  scientists in the world. The
second and third-rate ones do their best but do not get very far.
There are also first-rate people who make very important discoveries which
are of capital importance for the development of the science. Then
there are genius like Galillo and Newton. Ettore Majorana was one of
these. Majorana had greater gifts that anyone else in the world;
unfortunately he lacked one quality which other men generally have:
plain common sense" (E.Fermi {from Cocconi memories)}

"E. Majorana was very critical to himself and other people.
He was permanently unhappy with himself.  He was a pessimist but had very acute sense of humor.
He was conditioned by complicated and absolutely nontrivial living rules.
...E. Majorana was quite rich and I  can not avoid thinking that his life might not have finished so tragically should he have been obliged to work for a living. For that reason and also because he did not like to publish the results of all investigations he had made,
Majorana contribution to physics is much less than it could be "(B. Pontecorvo \cite{PontecorvoMaj}).

In conclusion I will  discuss briefly  the content  of the Majorana paper
"Symmetrical theory of electron and positron" \cite{EMajorana}.

E. Majorana was not satisfied with the existing at that time theory of electrons and
positrons in which positrons were  considered as holes in the Dirac sea
of the states of electrons with negative energies. He wanted to
formulate the symmetrical theory in which there is no notion of
states with negative energies.

Let us consider the Dirac equation for a complex field $\psi(x)$
\begin{equation}\label{A1}
(i \gamma^{\alpha}\partial_{\alpha} -m)~\psi(x)=0,
\end{equation}
where $m$ is the mass of the particles-quanta of the field.
The conjugated field
\begin{equation}\label{A2}
\psi^{c}(x)=C \bar\psi^{T}(x),
\end{equation}
($C$ is the matrix of the charge conjugation) obviously satisfies the same equation
\begin{equation}\label{A3}
(i \gamma^{\alpha}\partial_{\alpha} -m)~\psi^{c}(x)=0.
\end{equation}
Let us present the field $\psi(x)$ in the form
\begin{equation}\label{A4}
\psi(x)=
=\frac{\chi_{1} + i \chi_{2}}{\sqrt{2}},
\end{equation}
where
\begin{equation}\label{A5}
\chi_{1}(x)=\frac{\psi(x) + \psi^{c}(x)}{\sqrt{2}};\quad\chi_{2}(x)=\frac{\psi(x) - \psi^{c}(x)}{\sqrt{2}i}.
\end{equation}
It is obvious from (\ref{A1}), (\ref{A3}) and (\ref{A5})
that the fields $\chi_{1,2}(x)$ satisfy the Dirac equations
\begin{equation}\label{A6}
(i \gamma^{\alpha}\partial_{\alpha} -m)~\chi_{1,2}(x)=0.
\end{equation}
The fields $\chi_{1,2}(x)$ satisfy also {\em additional (Majorana) conditions} \begin{equation}\label{A7}
\chi^{c}_{1,2}(x)=\chi_{1,2}(x).
\end{equation}
Majorana used the representation in which $\gamma^{\alpha}$ are
imaginary matrices (Majorana representation). In this representation
$\psi^{c}(x)=\psi^{*}(x)$ and $\chi_{1}(x)$ and $\chi_{2}(x)$ are
real and imaginary parts of the field $\psi(x)$.

{\em Majorana build quantum field theory of the fields $\chi_{1,2}(x)$.} First of all it is easy to show that there are no
electromagnetic currents for the fields $\chi_{1,2}(x)$. In fact, taking into account (\ref{A6}), we have
\begin{equation}\label{A8}
j_{i}^{\alpha}(x) = \bar\chi_{i}(x)\gamma^{\alpha}\chi_{i}(x)=-\chi_{i}^{T}(x)(\gamma^{\alpha})^{T}\bar\chi_{i}(x)^{T}
=-\bar\chi_{i}(x)\gamma^{\alpha}\chi_{i}(x)=0;~~(i=1,2)
\end{equation}
Therefore,  $\chi_{1,2}(x)$ are fields of particles with electric charge and magnetic moment equal to zero.

For the operator  of the energy and momentum  Majorana obtained the
following expressions
\begin{equation}\label{A9}
P^{i}_{\alpha}=\int \sum_{r}p_{\alpha}(a_{r}^{i}(p))^{\dagger}~a^{i}_{
r}(p)~d^{3}p \quad (i=1,2).
\end{equation}
where operators $a^{i}_{r}(p)$ and $(a_{r}^{i}(p))^{\dagger}$ satisfy usual
anticommutation relations.

Thus, $(a_{r}^{i}(p))^{\dagger}$ ( $a^{i}_{r}(p)$) is the operator of the
creation (absorption) of a particle with momentum $p$ and helicity
$r$. There are no states with negative energies and quanta of the
fields $\chi_{1,2}(x)$ are neutral particles (which  are identical
to their antiparticles).

In the case of the complex field $\psi(x)= \frac{\chi_{1} + i\chi_{2} }{\sqrt{2}}$ the current
$j^{i}_{\alpha}(x) = \bar\psi_{i}(x)\gamma^{\alpha}\psi_{i}(x)$
is different from zero. After quantization Majorana came to  symmetrical theory of particles and antiparticles
with operators of total momentum and total charge given by the following expressions
\begin{equation}\label{A10}
P^{\alpha}=\int \sum_{r}p^{\alpha}[c^{\dagger}_{r}(p)c_{r}(p)+
d^{\dagger}_{r}(p)~d_{r}(p)]  d^{3}p
\end{equation}
\begin{equation}\label{A11}
Q=e\int \sum_{r}[c^{\dagger}_{r}(p)c_{r}(p) -
d^{\dagger}_{r}(p)~d_{r}(p)]  d^{3}p
\end{equation}
Here $c^{\dagger}_{ r}(p)(c_{ r}(p))$ is the operator of the
creation (absorption) of particle with charge $e$, momentum $p$ and
helicity $r$ and $d^{\dagger}_{ r}(p)(d_{ r}(p))$ is the operator
of the creation (absorption) of antiparticle with with charge $-e$,
momentum $p$ and helicity $r$. Correspondingly,
\begin{equation}\label{A12}
|p\rangle_{a} =c^{\dagger}_{r}(p)|0\rangle, \quad
|p\rangle_{\bar a} =d^{\dagger}_{r}(p)|0\rangle
\end{equation}
are states of particle with charge $e$, helicity $r$ and mass $m$ and antiparticle with charge $-e$, helicity $r$ and the same mass $m$.

Majorana wrote in the  paper \cite{EMajorana}: "A
generalization of Jordan-Wigner quantization method allows not only
to give symmetrical form to the electron-positron theory but also to
construct an essentially new theory for particles without electric
charge (neutrons and hypothetical neutrinos)". And further in the
paper: "Although it is perhaps not possible now to ask experiment
to choose between the new theory and that in which the Dirac
equations are simply extended to neutral particles, one should keep
in mind that the new theory is introducing in the unexplored field a
smaller number of hypothetical entities".

Soon after the Majorana paper   Racah \cite{Racah} and Furry \cite{Furry} proposed the methods which
could allow to test whether neutrino is Majorana or Dirac particle.
The so-called Racah chain of reactions
\begin{equation}\label{A13}
(A,Z) \to (A,Z+1) + e^{-}+\nu,\quad\nu+(A',Z')\to (A',Z'+1) +e^{-}
\end{equation}
is allowed in the case of the Majorana neutrino and is forbidden in
the case of the Dirac neutrino. Of course, in 1937 Racah could not
know that even in the case of the Majorana neutrino the chain
(\ref{A13}) is strongly suppressed due to neutrino helicity.

In 1938 Furry  considered neutrinoless double $\beta$-decay of nuclei
\begin{equation}\label{13b}
(A,Z) \to (A,Z+2) + e^{-}+ e^{-}
\end{equation}
induced by the Racah chain with virtual
neutrinos.

\end{document}